\documentclass[pre,showpacs]{revtex4-1}
\usepackage{amssymb,amsmath}
\usepackage{natbib}
\usepackage{graphicx}
\usepackage{color}

 \newcommand{\Mmat}[1]{\mathbb{#1}}

\def \ie {i.e.}

\defcitealias{PublicationI}{Publication I}

\begin{document}

\title{Hydrodynamic Interactions between Two Forced Objects
of Arbitrary Shape: II Relative Translation}

\author{Tomer Goldfriend}
\email{goldfriend@tau.ac.il}
\affiliation{Raymond \& Beverly Sackler School of Physics and Astronomy, Tel Aviv
University, Tel Aviv 69978, Israel}

\author{Haim Diamant} 
\email{hdiamant@tau.ac.il} 
\affiliation{Raymond \& Beverly Sackler School of Chemistry, Tel Aviv
University, Tel Aviv 69978, Israel}

\author{Thomas A.\ Witten} 
\email{t-witten@uchicago.edu}
\affiliation{Department of Physics and James Franck Institute,
University of Chicago, Chicago, Illinois 60637, USA}

\date{\today}

\begin{abstract}
We study the relative translation of two arbitrarily shaped objects,
caused by their hydrodynamic interaction as they are forced through a viscous fluid
in the limit of zero Reynolds number.
It is well known that in the case of two rigid spheres in an unbounded fluid,
the hydrodynamic interaction does not produce relative
translation.
More generally such an effective pair-interaction vanishes 
in configurations with spatial inversion symmetry; for example,
an enantiomorphic pair in mirror image positions has no relative translation.
We show that the breaking of inversion symmetry by boundaries of the system accounts
for the interactions between two spheres in
confined geometries, as observed in experiments.  
The same general principle also provides new 
predictions for interactions in other object configurations near obstacles.
We examine the time-dependent relative translation of two self-aligning objects,
extending the numerical analysis of our preceding publication
[Goldfriend, Diamant and Witten, Phys.~Fluids {\bf 27}, 123303 (2015)]. 
The interplay between the orientational interaction and the translational one,
in most cases, leads over time to repulsion between the two objects.
The repulsion is qualitatively different for self-aligning objects compared 
to the more symmetric case of uniform prolate spheroids. 
The separation between the two objects increases with time $t$ as $t^{1/3}$
in the former case, and more strongly, as $t$, in the latter.
\end{abstract}


\maketitle

\section{Introduction}
\label{sec:intro}

Hydrodynamic interactions are crucial for the
dynamics of colloidal dispersions~\cite{Happel&Brenner,Russel}.
These flow-mediated interactions are characterized by 
a long-ranged $R^{-1}$ decay with distance $R$.
The effect of hydrodynamic interactions is particularly strong 
in the case of driven systems, where an external force acts on each constituent object.
This effect is apparent already at the 
level of a single object, where the combination of driving and hydrodynamics 
generally leads to rotation-translation coupling~\cite{Happel&Brenner}.
At the level of a forced pair of objects the hydrodynamic interaction gives rise
to rich behavior, as presented in the preceding article~\cite{PublicationI} 
(referred to hereafter as Publication I) and the present one. 
On the collective level of driven suspensions, the long-ranged and strong hydrodynamic
interactions may create large-scale dynamical structures, as in colloid sedimentation~\cite{Ramaswamy2001}.

The hydrodynamic interaction between two forced symmetric objects,
e.g., spheres, has been explored extensively
in the middle of the last century; see Ref.~\cite{Happel&Brenner}
and references therein. In the limit of zero Reynolds number,
the hydrodynamic interaction between two identical sedimenting spheres,
isolated in an unbounded fluid, does not bring about any relative translation, i.e., the
spheres neither reduce nor increase their mutual distance while settling through
the fluid (they do not rotate around each other either). 
This remarkable result can be related to the time-reversal symmetry of
the Stokes equations governing the flow field~\cite{Ramaswamy2001}.
 
On the other hand, the vanishing relative velocity in the case of two spheres is 
readily violated by changing the system's geometry.
For example, Squires and Brenner~\cite{Squires&Brenner2000} pointed out
that, when forced away from a nearby wall, two spheres do develop relative velocity,
making them approach one another.
Such long-ranged attraction 
between two like-charged spheres in the presence
of a similarly charged wall was observed in optical tweezers experiments by
Larsen and Grier~\cite{Larsen&Grier1997}.
These apparent interactions do not originate from any direct,
e.g., electrostatic or van der Waals interaction, but from the velocity fields generated by the objects  
(and sometimes are referred to as ``hydrodynamic pseudo-potentials''~\cite{Squires2001}).
Another example of an attraction-like behavior, mediated 
by Stokes flow, appears in the motion
of two spheres driven along an optical vortex trap~\cite{Sokolov_etal2011}.    
The effects of such interactions can show up in experiments not only as pair-attractions
but also in more complex phenomena, such as collective phonon-like excitations in
driven object arrays~\cite{Nagar&Roichman2014,Beatus_etal2012}.

These previous studies of apparent interactions originating in hydrodynamic coupling were {\em ad hoc},
treating specific experimental scenarios.
In this article we address two more general questions: ({\it i}) At zero Reynolds number,
what are the geometrical configurations for which relative translation between
two objects necessarily vanishes?  
({\it ii}) In cases where it does not vanish, what are the consequences for
the long-time trajectories of the two objects?
Looking for properties of general applicability,
we consider arbitrarily shaped objects and do not restrict ourselves
to a specific geometry. 
Symmetry considerations have been successfully invoked in the past
for various hydrodynamic problems at zero Reynolds number, e.g.,
the motion of objects in shear flow~\cite{Bretherton1962,Makino&Doi2005},
or Purcell's theorem for swimmers~\cite{Purcell1977}.
Similarly, we seek general laws, derivable from symmetry arguments,
concerning the relative translation of driven object pairs.   

In the inertia-less regime the flow and the velocities of suspended objects at
a given moment are proportional to the external forces acting at that moment.  
Consequently, the motion of two interacting objects can be expressed by a grand
pair-mobility matrix~\cite{BrennerII,Brenner&Oneill1972,Kim&Karrila,PublicationI}. 
Earlier works focused on the mobility (or inversely, hydrodynamic resistance)
of objects in various simple geometries, such as a pair of spheres or 
spheroids~\cite{Goldman_etal1966,Wakiya1965,Felderhof1977,Jeffrey&Onishi1984,Liao&Krueger1980,Kim1985,Kim1986}.
In addition, several numerical techniques were developed to study dispersions of arbitrarily shaped
colloids~\cite{Karrila_etal1989,Cong&Thien1989,Carrasco&Torre1999,Kutteh2010,Cichocki_etal1994}.
In \citetalias{PublicationI}, we have 
studied general properties of the hydrodynamic interaction
and considered its effect on orientational dynamics.
In the present article, we extend this study, focusing on translational motion
of object pairs.

The motivation of Publication I was to understand the role 
of hydrodynamic interactions between {\em self-aligning objects}~\footnote{In 
\citetalias{PublicationI} they were referred to as ``axially alignbale'' objects.}.
An object is self-aligning if,
when subjected to an external unidirectional force (as in sedimentation),
it achieves terminal alignment between a specific eigen-direction
affixed to the object and the force direction, 
owing to a translation--rotation coupling in its mobility~\cite{Krapf_etal2009,Moths&Witten2013,Moths&Witten2013b}.
(See also Sec.\ref{ssec:single} below.)
We focused on self-aligning objects of irregular shape, which have a richer response as they also rotate
with a constant angular velocity about the aligning direction. 
To this end, we explored the pair-mobility of  
two identical, arbitrarily shaped objects, which are arranged in the same orientation.
Based on general considerations, 
and utilizing the system's symmetry under exchange of objects, we proved that self-aligning objects
undergo relative rotation, as well as relative translation, when forced through an unbounded fluid.
The leading effective interaction is dipolar,
scaling as $R^{-2}$ with the mutual distance $R$ between the objects. 
In addition, we used a numerical integration scheme 
to study the effect of these pair-interactions over time.
We found that the majority of our examples, comprising 
pairs of randomly constructed, self-aligning stokeslet objects, showed a repulsive-like behavior,
where the objects move away from each other in time.

These two key results, concerning the instantaneous and long-times interactions,
have led to the present work, which extends the analysis along two separate directions:
(a) In Sec.~\ref{sec:Instantaneous} we continue to study  
the instantaneous response of object pairs, i.e., the rigorous properties of the pair-mobility matrix.
We provide examples for configurations with spatial inversion symmetry, where the relative translation vanishes, as well as 
simple geometries, for which this symmetry is broken. 
(b) In Sec.~\ref{sec:alignable}
we return to examine in more detail the repulsive trend in the far-field time evolution
of self-aligning object pairs, providing a quantitative explanation of the phenomenon.
We further compare it to the time evolution of two non-alignable objects (uniform prolate spheroids),
and point out the qualitative difference between the two cases.  
Finally, the implications of our results, from theoretical and experimental points of view,
are discussed in Sec~\ref{sec:discussion}.

\section{Instantaneous Response}
\label{sec:Instantaneous}
\subsection{Pair Mobility Matrix}
\label{sec:PairMobility}

We consider a system of two rigid objects, $a$ and $b$, with typical
size $l$, subject to external forces and torques $\vec{F}^a$,
$\vec{F}^b$ and $\vec{\tau}^a$, $\vec{\tau}^b$ in a fluid of viscosity $\eta$.
The geometry of the system, i.e., the shape of each of the objects and
the fluid boundaries, are arbitrary.
Each of the objects is designated with an origin about which its linear velocity and torque
are measured. 
In the regime of zero Reynolds number (also known as Stokes flow, creeping flow or inertia-less flow),
the objects respond with instantaneous linear and angular velocities
through a symmetric, positive-definite $12 \times 12$
\emph{pair-mobility matrix}~\cite{Happel&Brenner,Brenner&Oneill1972,PublicationI}, 
\begin{equation}
   \begin{pmatrix}
  \vec{V}^a \\
	\vec{\omega}^a l\\
  \vec{V}^b \\
	\vec{\omega}^b l
 \end{pmatrix} =
\frac{1}{\eta l}
\begin{pmatrix}
  \Mmat{A}^{aa}	& (\Mmat{T}^{aa})^T	& \Mmat{A}^{ab} & (\Mmat{T}^{ba})^T\\
  \Mmat{T}^{aa} & \Mmat{S}^{aa}     & \Mmat{T}^{ab} & \Mmat{S}^{ab} \\
	(\Mmat{A}^{ab})^T	& (\Mmat{T}^{ab})^T	& \Mmat{A}^{bb} & (\Mmat{T}^{bb})^T\\
  \Mmat{T}^{ba} & (\Mmat{S}^{ab})^T     & \Mmat{T}^{bb} & \Mmat{S}^{bb} 
 \end{pmatrix}
   \begin{pmatrix}
  \vec{F}^a \\
	\vec{\tau}^a /l \\
  \vec{F}^b \\
	\vec{\tau}^b /l
 \end{pmatrix}.
\label{eq:GrandPair}
\end{equation}
The dimensionless blocks of the matrix defined in Eq.~\eqref{eq:GrandPair} depend on the whole geometry of the system 
(the shapes of the objects, their mutual configuration, as well as the shapes of the surrounding boundaries),
the boundary conditions at all the surfaces, and the choice of objects' origins.  
The transformation between matrices which differ in the choice of objects' origins can be found in 
Appendix A of \citetalias{PublicationI}. Hereafter we normalize the viscosity such that $\eta l=1$.

In this work we focus on the translational dynamics of two forced objects 
(without external torques),
which is given by a $6\times 6$ sub-matrix of the pair-mobility,
indicated hereafter by $\Mmat{A}$,
\begin{equation}
   \begin{pmatrix}
  \vec{V}^a \\
  \vec{V}^b 
 \end{pmatrix} =
 \begin{pmatrix}
  \Mmat{A}^{aa}	& \Mmat{A}^{ab}\\
  \Mmat{A}^{ba} & \Mmat{A}^{bb}
 \end{pmatrix}
   \begin{pmatrix}
  \vec{F}^a \\
  \vec{F}^b 
 \end{pmatrix}.
\end{equation}
The diagonal blocks, $\Mmat{A}^{aa}$ and $\Mmat{A}^{bb}$, correspond
to the response of an object to a force on itself, in the presence of the other object.  
The off-diagonal blocks, $\Mmat{A}^{ab}$ and $\Mmat{A}^{ba}$, correspond to the hydrodynamic
interaction, i.e., the response of one object to a force on the other.
The properties of the full pair-mobility matrix in Eq.~\eqref{eq:GrandPair} imply
that $\Mmat{A}$ is positive definite and symmetric~\cite{Condiff&Dahler1966,Happel&Brenner,Landau&Lifshitz}.
The matrix $\Mmat{A}$ has additional symmetries, as discussed in the next section. 

\subsection{Symmetrical Pair Configurations}

Brenner~\cite{BrennerII} (see also Ref.~\cite{Happel&Brenner}, Sec.~5.5)
characterized the properties of the self-mobility matrix for individual,
symmetric objects. Given a symmetry of the object's shape, 
he deduced which of the hydrodynamic responses vanish. For example, 
if an object has a plane of reflection symmetry, forcing it in the direction perpendicular
to that plane cannot lead to translation parallel to the plane.
For instance, forcing a spheroid along its major axis does not induce translation along
its two other principal axes.

As a first step,
we consider the transformation of the pair-mobility matrix under various operations--- 
spatial proper and improper rotations (rotations combined with reflections),
and exchange of objects.
We note that the positions of the objects' origins, i.e., the external forcing points,
are an essential part of the system geometry. 
In this section we will restrict our treatment to cases where the forcing point is at 
the geometric centroid of the object, that is, the center of mass if the object 
has a uniform mass density. Consider the transformation between two pair-mobility matrices which differ by 
a rigid rotation. 
Given the rotation matrix $\Mmat{R}$ between a given configuration and the rotated one,
the blocks in $\Mmat{A}$, being all tensors, transform accordingly:
$\Mmat{A}^{aa}\rightarrow\Mmat{R}\Mmat{A}^{aa}\Mmat{R}^T$,
$\Mmat{A}^{ab}\rightarrow\Mmat{R}\Mmat{A}^{ab}\Mmat{R}^T$
and
$\Mmat{A}^{bb}\rightarrow\Mmat{R}\Mmat{A}^{bb}\Mmat{R}^T$.
The same law applies to the transformation between systems which differ by a
rigid improper rotation with improper rotation matrix $\bar{\Mmat{R}}$.
We note here that the twist matrices $\Mmat{T}$, the blocks
in Eq.~\eqref{eq:GrandPair} which relate 
forces to angular velocities, are pseudo-tensors; thus, their transformation under
improper rotation includes a change of sign,
 $\Mmat{T}^{aa}\rightarrow-\bar{\Mmat{R}}\Mmat{T}^{aa}\bar{\Mmat{R}}^T$.
Since the pair-mobility matrix inherently refers to two distinguishable
objects, $a$ and $b$, we should also consider its transformation under
interchanging the objects' labeling, $a\leftrightarrow b$.
This transformation corresponds to interchanging the blocks
$\Mmat{A}^{aa}\leftrightarrow\Mmat{A}^{bb}$ and
$\Mmat{A}^{ab}\leftrightarrow\Mmat{A}^{ba}$;
or, in matrix form, 
\begin{equation}
 \Mmat{A} \rightarrow
 \Mmat{E}  \Mmat{A}   \Mmat{E}^{-1} ,
\label{eq:switch}
\end{equation}
where $\Mmat{E}$ is a $6\times 6$ 
matrix which interchanges the objects,
\begin{equation}
\Mmat{E}=
 \begin{pmatrix}
  0	& \Mmat{I}_{3\times3}\\
  \Mmat{I}_{3\times3} & 0
 \end{pmatrix},
\label{eq:Emat}
\end{equation}
and $\Mmat{I}_{3\times3}$ is the  $3\times 3$ identity matrix.

As mentioned in Sec.~\ref{sec:intro}, at zero Reynolds number, sedimentation of two identical rigid spheres
in an unbounded fluid does not induce relative motion of the two spheres.
In contrast, we showed in Publication I that two identical, arbitrarily shaped objects in an unbounded fluid,
under the same forcing, may attract or repel each other.
In addition, using the symmetry of such a system under exchange of objects,
we found the leading (dipolar) order of this effective interaction at large mutual separation.
We now examine the hydrodynamic interaction for several symmetric configurations.
In this part of the article we focus on what symmetry has to tell us,
and do not yet anticipate when its consequences are useful.
To demonstrate the resulting principles, therefore, we allow ourselves to examine specially designed configurations.
Examples for the usefulness of these principles will be given later on.

\subsubsection{Two Enantiomers in an Unbounded Fluid}
\label{sec:enantiomers}

As the first example, we examine a system possessing inversion symmetry.
The spatial inversion of the Stokes equations under reversing time has
been shown to imply fundamental consequences concerning the dynamics of rigid objects.
For example, it was used previously to deduce generic properties of shear flow response, in the cases of 
a single spheroid~\cite{Bretherton1962} and an enantiomeric non-interacting pair~\cite{Makino&Doi2005}.
The fact that the instantaneous response, under the same forcing,
of pair-configurations with spatial inversion symmetry does not induce relative translation
may have already been given elsewhere;
yet, we are not aware of works which give  a rigorous derivation of it.
Accordingly, we review this basic result below from matrix transformations and time-reversibility
points of view.   

\begin{figure}
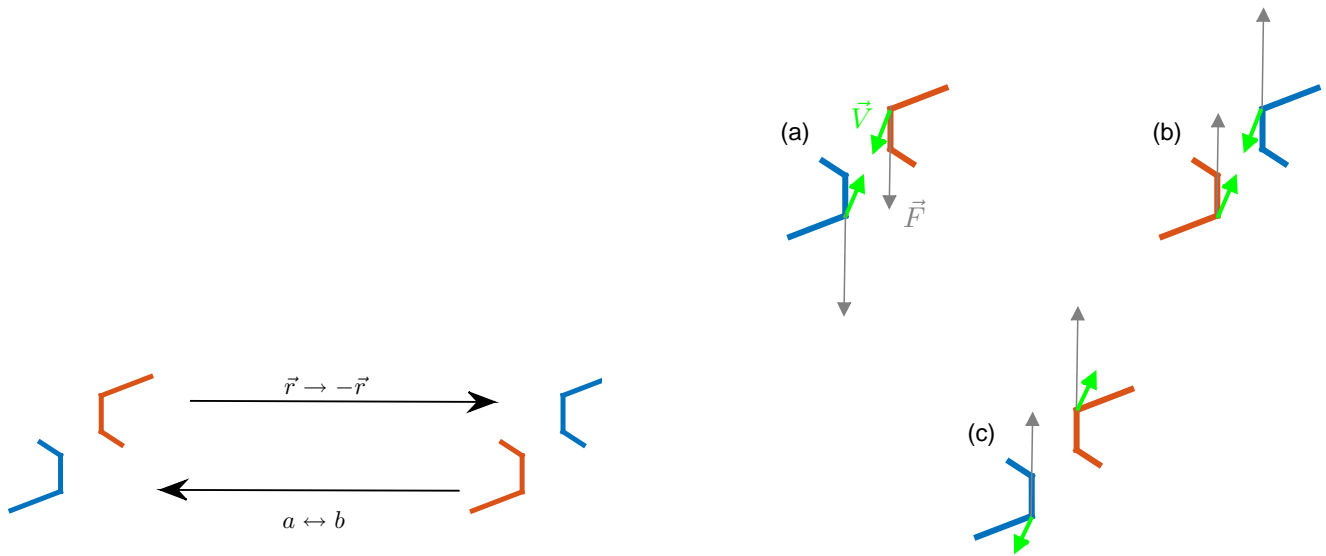

\centerline{\resizebox{0.45\textwidth}{!}{\includegraphics{fig1a.eps}}
\hspace{2cm}
\resizebox{0.4\textwidth}{!}{\includegraphics{fig1b.eps}}}
\caption[]{Schematic description of the arguments presented in Sec.~\ref{sec:enantiomers}
for the vanishing instantaneous relative motion between two forced enantiomers.
The left panel describes how the mirror symmetry of the system is expressed in terms of
spatial inversion combined with interchanging the objects.
The right panel demonstrates the arguments based on the
symmetry of the Stokes equations under inversion of time and forces,
where
gray (thin) and green (thick) arrows indicate forces and velocities respectively.
Configuration (b) is the spatial inversion of configuration (a) whereas 
configuration (c) is the response of configuration (a) under the opposite forcing.}
\label{fig:Enantiomers}
\end{figure}

Consider a system of two enantiomers in an unbounded fluid as depicted in 
the left panel of Fig.~\ref{fig:Enantiomers}. 
We choose the mutual orientation of the objects such that one object is the 
mirror image of the other; hence, the system's geometry is invariant under 
spatial inversion, $\vec{r}\to-\vec{r}$.
This symmetry implies that the pair-mobility matrix is invariant
under two consecutive operations (see
Fig.~\ref{fig:Enantiomers}): spatial inversion followed by exchange of objects' labels,
or, in matrix form, $\Mmat{A}=\Mmat{E}(-\Mmat{I}_{6\times 6})\Mmat{A}(-\Mmat{I}_{6\times 6})\Mmat{E}^{-1}$,
where $\Mmat{E}$ has been defined in Eq.~\eqref{eq:Emat}. 
This last equality yields $\Mmat{A}^{aa}=\Mmat{A}^{bb}=\Mmat{A}^{\text{self}}$
and $\Mmat{A}^{ab}=\Mmat{A}^{ba}=\Mmat{A}^{\text{coupling}}$, i.e., 
\begin{equation}
\text{Enantiomeric pair:}\ \ \ 
\Mmat{A}=\begin{pmatrix}
  \Mmat{A}^{\text{self}}	& \Mmat{A}^{\text{coupling}}\\
  \Mmat{A}^{\text{coupling}} & \Mmat{A}^{\text{self}}
 \end{pmatrix},
\label{eq:enantiomers}
\end{equation}
which applies in fact for any inversion-symmetric situation.
This form of the pair-mobility matrix implies that, under the same force $\vec{F}$,
the two objects will develop identical velocities,
$\vec{V}^a=\vec{V}^b=(\Mmat{A}^{\text{self}}+\Mmat{A}^{\text{coupling}})\cdot\vec{F}$.
Thus: the \emph{instantaneous} response under
the \emph{same forcing} of a pair, whose configuration possesses an inversion symmetry, does not
include relative translation.
As noted above, a system of two sedimenting spheres
is a particular example of this general result.  

The vanishing relative motion in a system
which is invariant under spatial inversion can be understood alternatively by the following argument.
Assume by negation that two enantiomers in an unbounded fluid develop a relative velocity
under the same forcing. Without loss of generality, let us take the case when the two objects get closer together;
see configuration (a) in the right panel of Fig.~\ref{fig:Enantiomers}.  
The mirror configuration of the system ($\vec{r}\to-\vec{r}$), depicted
in configuration (b), implies that the two also get closer when reversing the forces. 
On the other hand, Stokes equations are invariant under inversion of time and forces;
hence, reversing the forces in configuration (a) should make the objects get further apart, as depicted in
configuration (c). Since (b) and (c) represent the same system, we reach a contradiction, and
deduce that the relative velocity between the two enantiomers must vanish.

To summarize, it is inversion symmetry that governs the vanishing relative velocity
between two forced objects at zero Reynolds number; hence, whenever this symmetry
is broken one should expect relative translation.

An important remark bears mentioning here. The instantaneous rotational response of the enantiomeric pair corresponds    
to two \emph{opposite rotations},
i.e., non-vanishing relative angular velocity. (See \citetalias{PublicationI}.)
With time, the opposite rotations will break the inversion symmetry, unless there are additional 
symmetries, for example, when the objects' shapes are isotropic, as in the case of two spheres.
While the example of two enantiomers whose mutual symmetry is only instantaneous may seem artificial,
the principle that it demonstrates is useful.
Driving forces will sometimes bring bodies close to a symmetric situation,
and then one will find especially simple motions that call for explanation.
In particular, two irregular objects can be aligned by the driving~\cite{Krapf_etal2009,Moths&Witten2013,Moths&Witten2013b}
and then we should be interested in the question of whether they
keep their distance or drift apart.
Moreover, not only a pair of spheres will preserve inversion symmetry over time.
Another example is two ellipsoids, or indeed a pair of any bodies of revolution with fore-aft
symmetry, whose axes are aligned on a plane perpendicular to the force; see Fig.~\ref{fig:WallRing}(a).
Without any calculation, we can assert that such two objects will maintain their relative 
position over time.
The additional symmetry of the system imposes relative rotation only about the axes of symmetry
($y$-axis in Fig.~\ref{fig:WallRing}(a)), which does not break the inversion symmetry.
Another example will be given in the Discussion.

\subsubsection{Configurations with One Reflection Plane and Exchange Symmetry}
\label{sec:WallRing}

\begin{figure}
\centerline{\resizebox{0.3\textwidth}{!}{\includegraphics{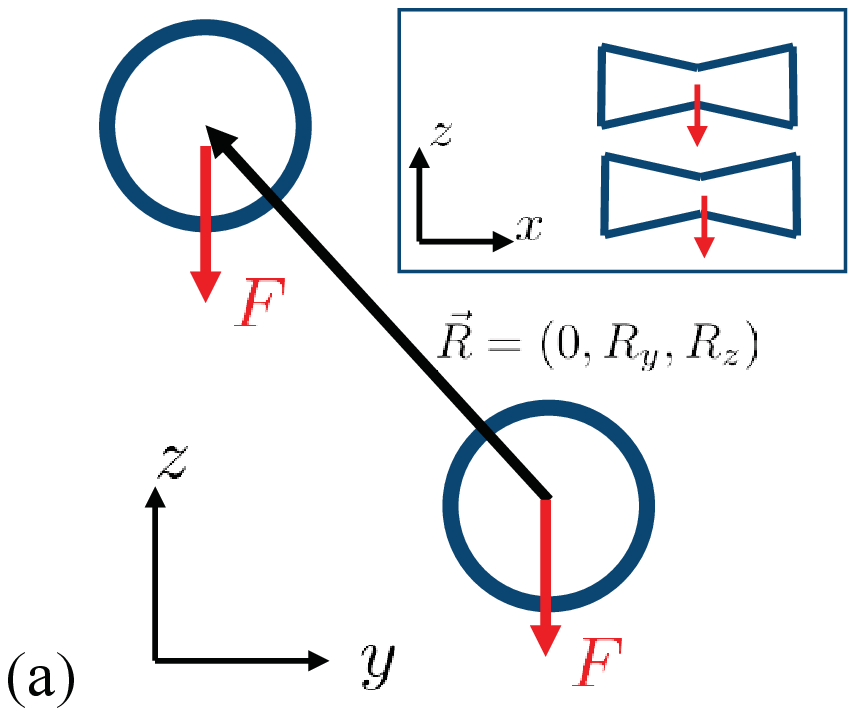}}
\hspace{0.5cm}
\resizebox{0.3\textwidth}{!}{\includegraphics{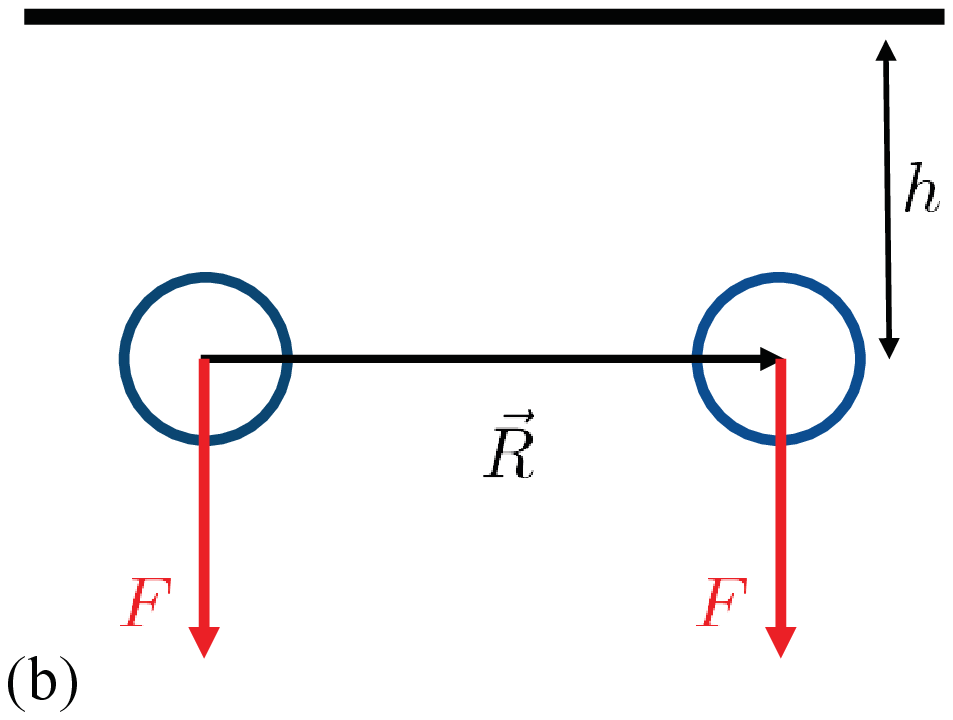}}
\resizebox{0.3\textwidth}{!}{\includegraphics{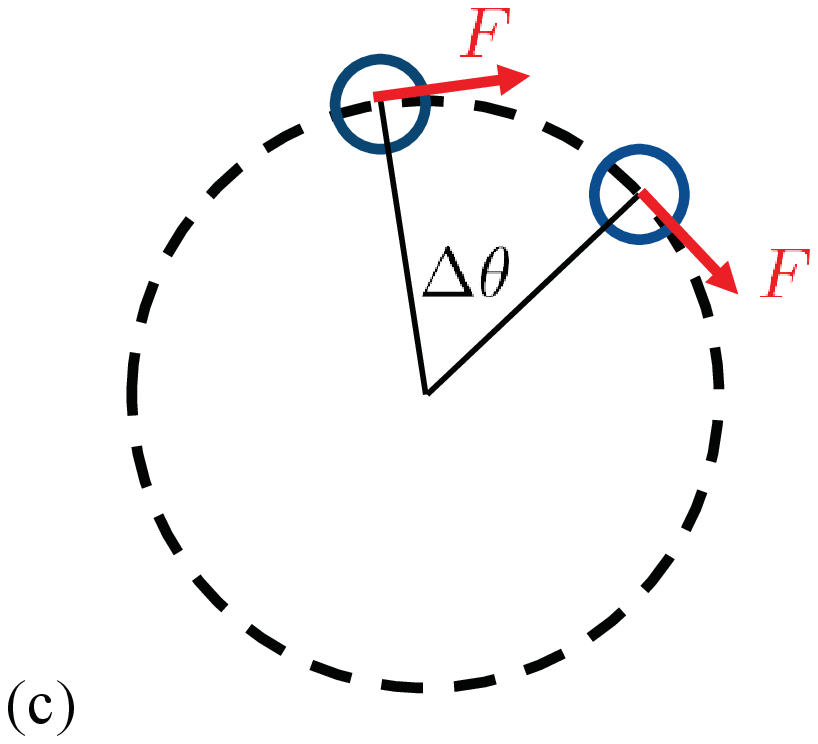}}}
\caption[]{
The three systems discussed in Secs.~\ref{sec:enantiomers} and~\ref{sec:WallRing}:
(a) Two identical axisymmetric objects,
with fore-aft symmetry, where the vector connecting their origins
is perpendicular to their symmetry axes.
(b) A pair of spheres near a wall. The spatial separation between the spheres is represented 
by the vector $\vec{R}$ which connects their origins.
(c) A pair of spheres driven along a ring.
The angular separation is indicated by $\Delta\theta$.
}
\label{fig:WallRing}
\end{figure}

We now turn to examples where the symmetry is broken by the confining boundaries.
The first system consists of two spheres, placed on a plane parallel to a wall; see 
Fig~\ref{fig:WallRing}(b).
This system was used in Ref.~\cite{Squires&Brenner2000} to
interpret the attraction between two like-charged spheres near a similarly charged wall, as
was observed in optical-tweezers experiments~\cite{Larsen&Grier1997}.
The geometry of the system is invariant under two successive operations: 
reflection about the symmetry plane of the two spheres and interchanging the objects.
We denote by $\parallel$ and $\perp$ the directions parallel and perpendicular, respectively, to
the mirror plane, i.e., perpendicular and parallel, respectively, to the wall itself. 
The reflection leaves vectors parallel to the mirror plane 
unchanged but reverses vectors normal to that plane. 
Using the transformation laws introduced above, we find that the blocks of
$\Mmat{A}$, which relate forces perpendicular to the wall 
and objects' velocities parallel to the wall, must 
satisfy $\Mmat{A}^{aa}_{\perp \parallel}=-\Mmat{A}^{bb}_{\perp \parallel}$ and
 $\Mmat{A}^{ab}_{ \perp \parallel}=-\Mmat{A}^{ba}_{ \perp\parallel}$.
These restrictions imply that, under forcing toward or away from 
the wall
(e.g., as the spheres are electrostatically repelled from the wall~\cite{Larsen&Grier1997}),
the objects respond in opposite directions in the plane parallel
to the wall,
$\vec{V}^a_{\perp}=(\Mmat{A}^{aa}_{\perp \parallel}+\Mmat{A}^{ab}_{\perp \parallel}) F_{\parallel}=-\vec{V}^b_{\perp}$.
Note that the symmetry of the system alone does not tell us whether the objects repel or attract under 
a given forcing direction. 
The analysis in Ref.~\cite{Squires&Brenner2000} showed that 
forcing away from the wall results in an apparent attraction between the pair,
in agreement with the experiment.
It should be stressed that, unlike specific calculations as in Ref.~\cite{Squires&Brenner2000},
our symmetry principle is neither restricted to spheres, nor to the limit of small objects,
nor to the limit of large separations.

The complete form of the pair-mobility matrix as a result of the geometrical
restrictions in this system is given in Appendix~\ref{app:Wall}, Eq.~\eqref{eq:FullWall},
along with the explicit expressions for point objects.
Another interesting conclusion, arising solely from the system's symmetry, is that,
since $\Mmat{A}^{aa}_{\perp \parallel}=-\Mmat{A}^{bb}_{\perp \parallel}$ and
 $\Mmat{A}^{ab}_{\perp \parallel}=-\Mmat{A}^{ba}_{ \perp \parallel}$, forcing the spheres 
{\em parallel} to the wall
will result in one sphere approaching the wall and the other moving away from it.

As a second example, let us consider the system addressed in Ref.~\cite{Sokolov_etal2011} ---
two spheres forced to move along a ring; see Fig.~\ref{fig:WallRing}(c).
The corresponding pair-mobility matrix can be written in polar coordinates $(\rho,\theta)$.
The system is invariant under two successive operations:
inversion of the $\theta$ coordinate, $\theta \to -\theta$, and interchanging the objects.
Hence, this system is similar to the previous one in the sense that it is symmetric
under objects exchange by inversion of one coordinate.
This leads to $\Mmat{A}^{aa}_{\rho \theta}=-\Mmat{A}^{bb}_{\rho \theta}$ and
 $\Mmat{A}^{ab}_{\rho \theta}=-\Mmat{A}^{ba}_{\rho \theta}$. 
Consequently, under the same forcing along the tangential direction,
the spheres respond with opposite velocities along the radial direction.
Sokolov et al. used a holographic optical vortex trap to study this system and
observed the radial symmetry breaking experimentally~\cite{Sokolov_etal2011}.
In addition, they found that this effect, combined with a confining radial potential,
results in overall attraction along the ring as the
system evolves in time.
Once again, the symmetry argument derived here is far more general than the specific limit studied theoretically
in Ref.~\cite{Sokolov_etal2011}.

We note that the results obtained above, regarding configurations with one reflection plane
of symmetry, should also be derivable using time-reversal arguments,
as was done in the case of an enantiomeric pair.

\section{Far-Field Dynamics of two Forced Objects in Unbounded Fluid}
\label{sec:alignable}

In the preceding section we considered the instantaneous response of 
a pair of objects given the symmetries of their configuration.
The analysis of this linear problem, derived from symmetry arguments,
is useful to determine the stability of a given state of the system.
However, it might be inapplicable to the time-dependent trajectories,
since a configuration symmetry at a given time can be broken by subsequent motion.
A well known example is the sedimentation in an unbounded fluid of two prolate spheroids, 
which start with their major axes parallel to the force~\cite{Wakiya1965,Kim1985,Kutteh2010}.
The initial inversion symmetry about the plane perpendicular to the force, which,
according to the discussion in Sec.~\ref{sec:Instantaneous}, precludes any
relative translation, soon breaks due to the rotation of each spheroid,
and relative velocity appears (see a more detailed analysis below).   
In general, since the instantaneous response depends on configuration,
the time-dependent trajectories of a driven pair is a non-linear, multi-variable problem.
We are compelled, therefore, to implement numerical integration for specific examples,
and try to identify general trends.
In this section we consider the time evolution of two objects under the same constant driving.
In particular, we provide further insight into the results reported in Publication I
concerning the combined effects of rotational and translational interactions.

\subsection{Dynamics of an isolated object}
\label{ssec:single}

Before describing the time evolution of object pairs, it is essential to introduce the different 
types of objects that we consider hereafter and describe their dynamics,
under a unidirectional force, at the single-object level. For more details on the various orientational behaviors
of single objects, see Refs.~\cite{Krapf_etal2009,Gonzalez_etal2004}.
In the absence of external torque, the linear and angular velocities of an object are given, respectively, 
by $\Mmat{A}_0\cdot\vec{F}$ and $\Mmat{T}_0\cdot\vec{F}$, where 
$\Mmat{A}_0$ and $\Mmat{T}_0$ are $3\times 3$ blocks of the object's self-mobility
matrix. These blocks depend on the shape of the object, its orientation, and the position of the
forcing point. 

We consider three types of objects: 
(a)~{\it Uniform prolate spheroid} --- a spheroid whose forcing point is located at its geometric centroid,
as in the case of spheroids with a uniform mass density.
For such an object $\Mmat{T}_0=0$; hence, it does not rotate, regardless of its orientation.
The translation direction of a uniform spheroid is in the plane spanned 
by its major axis and the forcing direction~\cite{Happel&Brenner}.
(b)~{\it Self-aligning prolate spheroid} --- a prolate spheroid whose forcing
point is displaced from the centroid along its major principal axis,
e.g., spheroids with a non-uniform mass density. These objects
have an antisymmetric $\Mmat{T}_{0}$ matrix.
For any initial orientation, a self-aligning spheroid rotates toward
a state where its major axis, and therefore its translation direction,
is aligned with the external force.
We use type (b) as a simple example, which can be treated analytically,
for self-aligning objects.
(c)~{\it Self-aligning object of irregular shape} --- an object whose $\Mmat{T}_0$ matrix
has only one real, non-zero eigenvalue. 
These objects reach an ultimate alignment between a specific eigen-direction
affixed to the object (the eigenvector corresponding to the real eigenvalue)
and the force, together with a uniform right- or left-handed rotation about it (according
to the sign of the real eigenvalue)~\cite{Krapf_etal2009,Moths&Witten2013,Moths&Witten2013b}.
As in Publication I we use the particularly simple construction of 
stokeslet objects --- a discrete set of small spheres, separated
by much larger, rigid distances, where each sphere is approximated by
a point force. Constructing them randomly, we avoid objects of pre-designed shapes.   
	
\subsection{Far-Field Equations}
\label{ssec:FarField}

Let us assume that there are no external torques on the objects,
such that we can choose their origins as their forcing points.
We consider the case in which the objects are subjected to the same forcing $\vec{F}$.
The mutual separation between the objects $a$ and $b$ is designated with
the vector $\vec{R}$, whose direction is defined from the origin of $a$ to the origin of $b$.
While time-integrating the equations it is essential to take into account the coupling
between translation and rotation. Thus, we must work with the complete pair-mobility
matrix, Eq.~\eqref{eq:GrandPair}, which gives
\begin{eqnarray}
\dot{\vec{R}}= \vec{V}^a-\vec{V}^b &=& \left(\Mmat{A}^{aa}+\Mmat{A}^{ab}
-\Mmat{A}^{bb}-\Mmat{A}^{ba}\right)
\cdot\vec{F} \\
\vec{\omega}^a&=& \left(\Mmat{T}^{aa}+\Mmat{T}^{ab}\right)\cdot\vec{F} \\
\vec{\omega}^b&=& \left(\Mmat{T}^{bb}+\Mmat{T}^{ba}\right)\cdot\vec{F}. 
\end{eqnarray}

A major simplification, from both the analytical and the numerical points of view, is to consider pairs with separation
much larger than the typical size of the individual object $l$,
and study the corresponding far-field
interaction. In \citetalias{PublicationI} we studied a system of
two arbitrarily shaped objects in an unbounded fluid, and derived the general form of the pair-mobility matrix up
to second order in $l/R$. The zeroth-order term corresponds to two non-interacting objects; 
The first-order term accounts for the advection of one object by 
the Oseen flow generated by the other, which results in a common translation of the pair
with velocity $\Mmat{G}(\vec{R})\cdot\vec{F} \sim R^{-1}$ and with no relative translation;
The second order term, $\sim\vec{\nabla}\Mmat{G}\sim R^{-2}$,
is the leading term which can give rise to relative
translation between the objects via hydrodynamic interactions.

The equations governing the objects' mutual separation and 
their rotations read
\begin{eqnarray}
\dot{\vec{R}} &=& \left[(\Mmat{A}_0^a-\Mmat{A}_0^b)+ 
(\Pi^a +\Pi^b) : \vec{\nabla}\Mmat{G}(\vec{R}) - \vec{\nabla}\Mmat{G}(\vec{R})^T: (\Pi^a +\Pi^b)^T  
\right]\cdot\vec{F} \label{eq:EOM1}\\
\vec{\omega}^a&=& \left[ \Mmat{T}_0^a +(\Psi^a-\frac{1}{2} \mathcal{E}): \vec{\nabla}\Mmat{G}(\vec{R})
\right]\cdot\vec{F} \label{eq:EOM2}\\
\vec{\omega}^b&=& \left[ \Mmat{T}_0^b -(\Psi^b-\frac{1}{2} \mathcal{E}): \vec{\nabla}\Mmat{G}(\vec{R})
\right]\cdot\vec{F} , 
\label{eq:EOM3}
\end{eqnarray}
where $\Mmat{A}_0$, $\Mmat{T}_0$ and $\Pi$, $\Psi$ are single-object-dependent tensors of rank 2 and 3, respectively, 
which depend on the individual object's shape and orientation,
and $\mathcal{E}$ is the Levi-Civita tensor.
In Publication I we introduced a tensor $\Phi$ with dimensions $6\times 3\times 3$;
here we separate it into its translational part, $\Pi$, and rotational part, $\Psi-\frac{1}{2}\mathcal{E}$,
each with dimensions $3\times 3\times 3$. 
The tensors $\Mmat{A}_0$ and $\Mmat{T}_0$ are the zeroth order blocks (the blocks in the self-mobility matrices),
which give the linear and angular velocities of a single object when it is subjected to external force.
The tensors $\Pi$ and $\Psi$ correspond to the linear- and angular-velocity 
responses to a flow gradient at the object's
origin. When these tensors are coupled with $\vec{\nabla}\Mmat{G}$
they construct second-order terms of the pair-mobility matrix, describing 
the direct hydrodynamic interaction between the objects.
The term which is proportional to $\mathcal{E}$ is also a part of the second-order term,
giving the rotation of one object
with the vorticity generated by forcing the other.
The tensors $\Mmat{A}_0$, $\Mmat{T}_0$ and $\Pi$  depend on the choice of objects'
origins; for the corresponding transformations see Appendices A and B in \citetalias{PublicationI}, or
Ref.~\cite{Kim&Karrila}.

\subsection{Transversal Repulsion under Constant Forcing}
\label{ssec:repulsion}

In Publication I we examined the effect of hydrodynamic interactions on the 
orientational evolution of two identical, self-aligning objects. 
Using numerical integration we followed the time-dependent trajectories of 
pairs of stokeslet objects under two types of driving--- 
a constant force and a rotating one.
The latter, {\em in the absence} of hydrodynamic interactions,
tends to synchronize each object with the rotating force~\cite{Moths&Witten2013,Moths&Witten2013b}. 
We noticed that in most (though not all) of the studied examples the two identical,
self-aligning objects effectively repelled each other when subjected to the same 
driving (see solid red curve in Fig.~\ref{fig:repulsion}(a)). 
(Counter-examples, such as limit-cycle trajectories, were observed as well.)
The increasing separation is transversal--- taking place within the plane perpendicular to the average force direction.
Below we explain the nature of this repulsion, focusing on the simpler case of constant forcing.

\subsubsection{Two self-aligning objects}
\label{sssec:selfaligning}

\begin{figure}
\centerline{\resizebox{0.5\textwidth}{!}{ \includegraphics[scale=10]{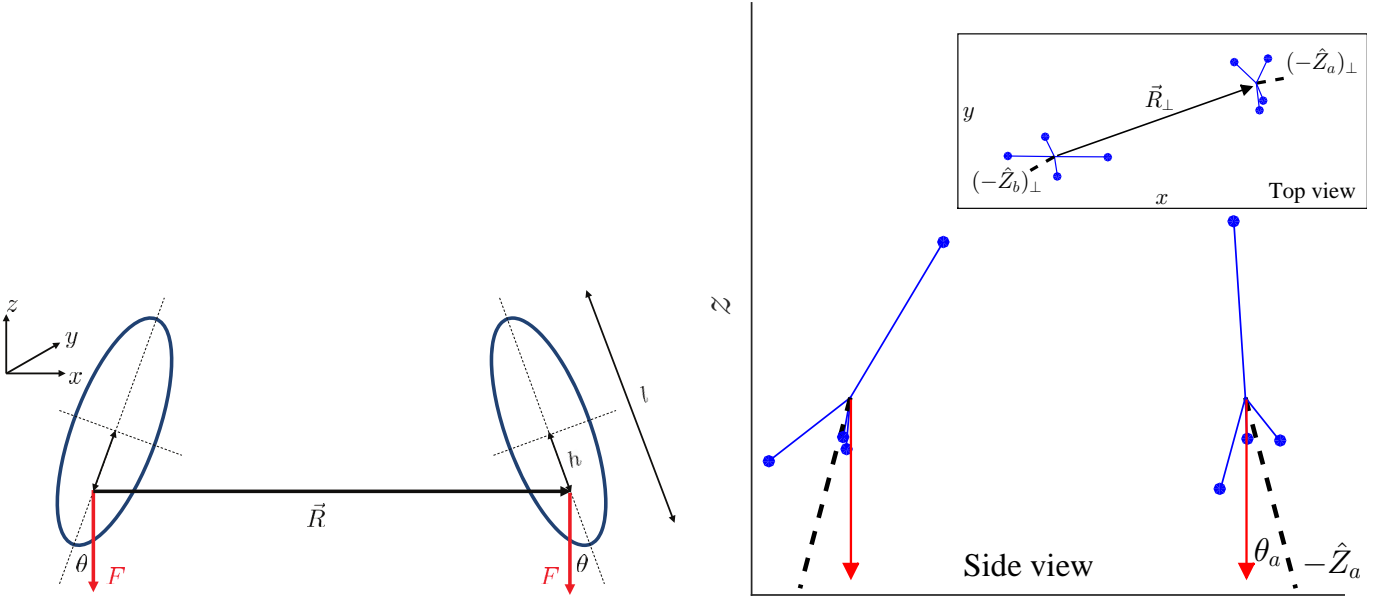}}
\hspace{0cm}
\resizebox{0.5\textwidth}{!}{
\includegraphics[height=10cm,clip=true]{fig3b.eps}
\llap{ \raisebox{6.5cm}{
\includegraphics[height=3.cm,clip=true]{fig3binset.eps}}}}}
\caption[]{Left panel: system of two self-aligning, prolate spheroids. The tilt angle $\theta$ is between
the force and the major axis of the object, and $h$ indicates the shifted position of the forcing point.
Right panel: system of two identical self-aligning objects made of 4 stokeslets.
In the absence of interaction the eigen-direction of each object $(-\hat{Z})$ would eventually 
align with the force which is along the $(-\hat{z})$ axis (not drawn), and rotate about it with
constant angular velocity. The hydrodynamic interaction tilts the objects by angles $\theta_a$ and 
$\theta_b$, while keeping a correlation between the transversal direction of the eigen-directions
and the separation vector. This effect results in repulsion
between the objects while they continuously rotate in the $xy-$plane.
Here we show a snapshot of this terminal evolution.
The main figure shows the projection of the system (the objects
and their eigen-directions) onto the $xz$-plane, whereas the inset shows the $xy-$plane projection,
together with the separation vector $\vec{R}$. }  
\label{fig:spheroids}
\end{figure}

Let us consider the time evolution of the following system, depicted in Fig.~\ref{fig:spheroids}:
two identical self-aligning spheroids, positioned initially along the $x$-axis,
and subjected to a constant force along the $(-\hat{z})$ direction.
Self-aligning spheroids are achieved by separating the center of forcing from their centroids, e.g.,
through a nonuniform mass density under gravity.
The configuration has an inversion symmetry about the $yz$-plane. 
It does not have an inversion symmetry about the $xy$-plane unless the spheroids are aligned along $\hat{z}$.
Thus, according to Sec.~\ref{sec:Instantaneous}, unless aligned,
they are expected to have an instantaneous relative velocity.
We denote by
$\theta$ the angle between the force and the major axis of each spheroid; $l$ indicates the
length of the major axis, and $h$ is the displacement of the forcing point from the centroid.
For given $h$ and $\theta$, the individual-object's tensors which appear in Eqs.~\eqref{eq:EOM1}--\eqref{eq:EOM3}
can be found from the known tensors for $h=0$ and $\theta=0$ (e.g. Ref.~\cite{Happel&Brenner}) 
by a change of origin and rotation transformation.
According to Sec.~\ref{ssec:single},
{\em in the absence} of interactions and $\theta\neq 0$, two uniform prolate spheroids will maintain
their relative tilt and glide away from each other with a constant velocity, whereas two self-aligning ones will do
the same but with velocity decreasing in time as they get aligned with the force.   
In order to examine the effect of hydrodynamic interactions we take the initial condition $\theta(t=0)=0$, 
for which relative translation vanishes in their absence.
Using the calculated individual-object's tensors for self-aligning spheroids, the equations of motion for the pair in the far-field regime $R \gg l$ and $\theta \ll 1$,
Eqs.~\eqref{eq:EOM1}--\eqref{eq:EOM3}, are then reduced
to the following simple form (recall that we set $\eta l=1$):
\begin{eqnarray}
\label{eq:ellipsEOM1} 
\dot{R} &=&  \left[ \alpha \theta   + \zeta \left( \frac{l}{R} \right)^2 \right] \frac{F}{8\pi}\\
l \dot{\theta} &=& \left[ -\lambda \theta   + \left( \frac{l}{R} \right)^2 \right] \frac{F}{8\pi}, 
\label{eq:ellipsEOM2}
\end{eqnarray}
where the dimensionless parameters $\alpha$, $\zeta$ and $\lambda$ 
(derivable from the single-object tensors) depend
on the spheroid's aspect ratio and $h/l$. (More details on the derivation of 
the above equations are given in Appendix~\ref{app:numerical}.) 
Using the conventions of Fig.~\ref{fig:spheroids}, $\alpha>0$ and $\lambda\geq 0$;
hence, positive $\theta$ implies increasing separation and decreasing tilt.
Differentiating Eq.~\eqref{eq:ellipsEOM1}, and substituting $\dot{\theta}$ from
Eq.~\eqref{eq:ellipsEOM2}, we obtain the equation for the separation alone,
\begin{equation}
\ddot{x}=-(\lambda+2\zeta x^{-3} ) \frac{F}{8\pi l} \dot{x}
+ (\alpha+\lambda\zeta )x^{-2} \left(\frac{F}{8\pi l}\right)^2,
\label{eq:x}
\end{equation}
where we have set $x\equiv R/l$.

The translational dynamics is dictated by ({\it i}\,) opposite mutual
glide of one object away from the other, and ({\it ii}\,) direct
hydrodynamic interaction which decays as $R^{-2}$.  (We neglect the
higher-order correction of the interaction $\sim \theta R^{-2}$.)  The
evolution of the tilt angle $\theta$ is governed by two competitive
effects--- the vorticity which increases it and the tendency of the
individual spheroid to align with the force.  The fact that the effect
of direct hydrodynamic interaction on the angular velocities is
independent of the object's shape is specific to configurations in
which $\vec{R}\perp\vec{F}$, regardless the object's
geometry~\setcounter{footnote}{10}\footnote{The object-dependent tensor $\Psi$, which appears
  in Eqs.~\eqref{eq:EOM2} and \eqref{eq:EOM3}, characterizes the
  response to a spatially symmetric flow gradient only (see Appendix B
  in \citetalias{PublicationI}).  When $\vec{R}\perp\vec{F}$, the flow
  gradients created by the force monopole on each object,
  $\pm\vec{\nabla}\Mmat{G}(\vec{R})\cdot\vec{F}$, are
  antisymmetric. Thus, the object-dependent terms vanish, and the
  angular velocities are solely affected by the vorticity $\sim
  \hat{R}\times\vec{F}/R^2$.}.  The effect of additional separation
along the force direction is discussed at the end of this section.

\begin{figure}
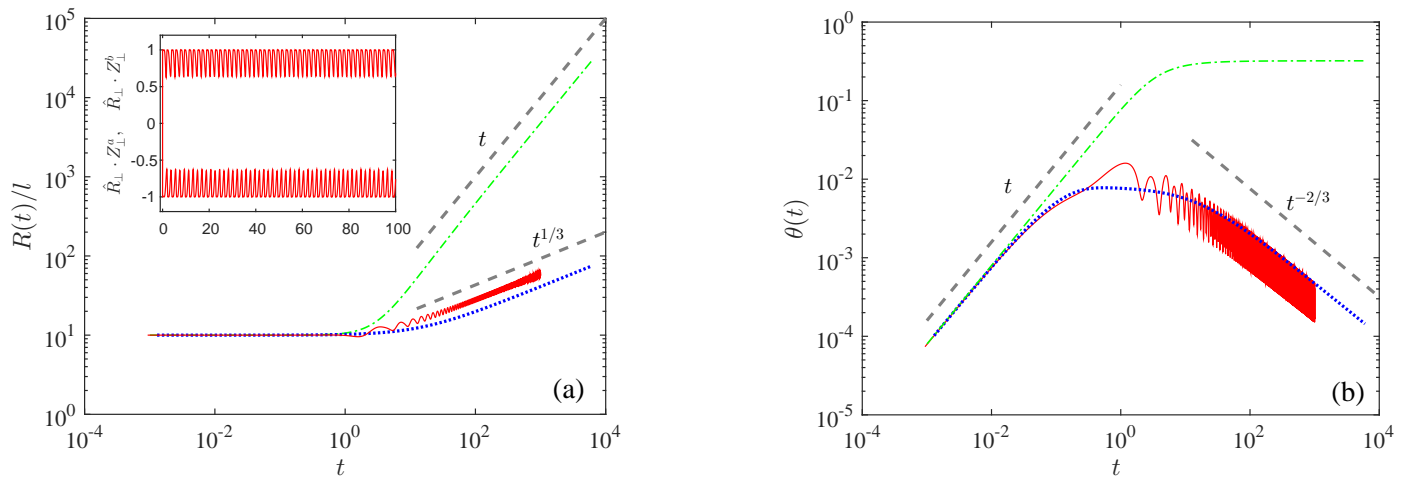

\centerline{\resizebox{0.45\textwidth}{!}{\includegraphics{fig4a.eps}}
\hspace{2cm}
\resizebox{0.45\textwidth}{!}{\includegraphics{fig4b.eps}}}
\caption[]{Representative trajectories for systems of two uniform
  spheroids (dash-dotted green curves), a pair of self-aligning
  spheroids (dotted blue curves), and two self-aligning stokeslets
  objects (solid red curves), together with the asymptotic behaviors
  at short and long times (gray dashed lines).  The results are
  presented in dimensionless units by setting the parameters of length
  and time to $l=1$ and $(\lambda F/8\pi l)^{-1}=0.1$, and the
  additional parameters to the values detailed in
  Appendix~\ref{app:numerical}. The distance between the objects is
  shown in panel (a), exhibiting the effective repulsion between the
  two objects. The inset demonstrates the correlation between the
  eigen-directions and the separation vector in the case of
  self-aligning irregular objects.  Panel (b) shows the evolution of
  the tilt angle (the solid red curve corresponds to only one of the
  objects). }
\label{fig:repulsion}
\end{figure}

The dotted blue curves in Fig.~\ref{fig:repulsion} show an example of
a numerical solution of
Eqs.~\eqref{eq:ellipsEOM1}--\eqref{eq:ellipsEOM2}. Initially $\theta$
increases linearly due to the vorticity term in
Eq.~\eqref{eq:ellipsEOM2}. After a typical time of $\sim (\lambda
F/8\pi l)^{-1}$ this increase is suppressed by the alignability of
each object, and at $t\to \infty$, the separation increases as
$t^{1/3}$ while $\theta$ decreases to 0 according to a $t^{-2/3}$ law.
These asymptotic laws can easily be inferred analytically.
The growth in mutual separation is a combination of a gliding term
($\alpha$) and an interaction term ($\zeta>0$). We note, however, that the
$t^{1/3}$ law arises from the alignability alone, 
whereas the direct interaction can only quantitatively affect the dynamics.
As seen from Eq.~\eqref{eq:x} in the limit of large $x$, the
equivalent system in classical mechanics is the damped equation
$\ddot{x} = -B \dot{x} + A x^{-2}$. At long times, $t\gg B^{-1}$,
acceleration is negligible and we are left with the equation
$\dot{x}=(A/B) x^{-2}$ for the velocity. This equation yields the
terminal $x\sim t^{1/3}$ law of the dotted blue curve in
Fig.~\ref{fig:repulsion}(a).

Next, we consider the repulsive time-dependent trajectories in the more general case of
two self-aligning objects of irregular shape.
We emphasize that this general case is expected to show a richer behavior;
this has been illustrated in Publication I
for stokeslets objects, where, for example, attractive-like behavior was observed as well.  
The repulsive trend reported in \citetalias{PublicationI} 
occurs in the majority ($\sim 80 \%$) of our several dozens examples comprising randomly constructed 4-stokeslets objects.
In addition, in the far-field limit, it is independent of the initial separation 
along the force direction, as well as the initial mutual orientation.
Here, we illustrate how the theoretical result derived for self-aligning spheroids is evident
also in the effective repulsion between two self-aligning objects of arbitrary shape.

Self-aligning objects of irregular shape exhibit complex dynamics already on the single-object level,
as they acquire ultimate rotation about their eigen-direction.
For example, their terminal translation direction is not
necessarily constant and might rotate about the forcing direction.
Here we consider two identical arbitrarily shaped objects, in which the
pair configuration has no spatial symmetry.  The coordinate space
includes the mutual separation $\vec{R}$ and the orientation variables
of each object, where we represent the latter with Euler-Rodriguez
4-parameters (or unit
quaternions)~\cite{PublicationI,Favro1960,Kutteh2010}.  The resulting
equation of motion for this coordinate space is a set of coupled
non-linear, first-order ODEs, which can be solved with conventional
techniques; see Appendix~\ref{app:numerical} for details of the
integration scheme.  The objects are initially separated along the
$x$-axis and aligned with the force (which, as before, is along the
negative $z$-axis).  These specific initial conditions are used to
emphasize the comparison with the pair of spheroids.  As opposed to
the spheroid pair, where the eigen-directions rotate only about the
$y$-axis and the separation unit-vector $\hat{R}$ is fixed to its
initial direction $\hat{x}$, here the former and the latter undergo a
complex, 3D rotational motion.

We follow the dynamics of each object's eigen-direction, which is
affixed to the object-reference-frame and denoted by $-\hat{Z}$.  For
each object we define a tilt angle $\cos\theta(t)\equiv\ \hat{Z}(t)
\cdot \hat{z}$, and an azimuthal correlation with the separation
vector $\cos\phi(t)\equiv\ -\hat{Z}_{\perp}(t)
\cdot\hat{R}_{\perp}(t)$.  A scheme of the configuration with the
relevant variables is depicted in Fig.~\ref{fig:spheroids}. The two 
objects can effectively glide away from each other, in resemblance to the
case of two self-aligning spheroids, if the eigen-directions are
tilted away from the separation direction, that is,
$\cos\phi^{\,a}(t)=1$ and $\cos\phi^b(t)=-1$; see inset in the right
panel of Fig.~\ref{fig:spheroids}.  The solid red curves in
Fig.~\ref{fig:repulsion}, which correspond to a representative
example, consisting of two identical objects, demonstrate that such
resemblance between the two cases does exist.  In particular, the solid
red curve in Fig.~\ref{fig:repulsion}(b) shows that $\theta(t)$ of
object $a$ follows the same trend as the dotted blue curve which
represents the simple example of self-aligning spheroids. The tilt
angle of object $b$, which is not shown, has a similar behavior.
The inset in Fig.~\ref{fig:repulsion}(a) shows the opposite correlations
between $\hat{Z}^{a}$, $\hat{Z}^{b}$ and $\vec{R}$. The gliding effect
results in an effective repulsion, $R(t)\propto t^{1/3}$, as can be
seen in Fig.~\ref{fig:repulsion}a. The direct interaction term in
Eq.~\eqref{eq:EOM1}--- proportional to $\Pi$ and decaying as
$R^{-2}$--- can also contribute to the $t^{1/3}$ trend.

The example of an arbitrarily shaped, self-aligning pair, presented in Fig.~\ref{fig:repulsion},
is a representative of a half-dozen other examples not shown here. These randomly generated examples
correspond to initial conditions, which involve also longitudinal separation and different mutual orientations.
The variance in the measured exponents is within a small numerical error, of order 5\%.
The analytically predicted power law was found to hold for {\it all} 
pairs of objects which drifted far apart in the simulations (80\% of the examples).
This implies that the $1/3$ exponent for the asymptotic repulsion
is most probably general for the class of self-aligning objects. 

\subsubsection{Two uniform prolate spheroids}
\label{sssec:uniform}

A pair of uniform spheroids exhibits quite different behavior from that 
of two self-aligning objects.
An individual uniform spheroid ($h=0$) does not rotate under forcing, and does not
translate in response to a flow gradient, i.e., the $\Mmat{T}_0$
matrices and $\Pi$ tensors in Eqs.~\eqref{eq:EOM1}--\eqref{eq:EOM3} 
vanish~\footnote{In \citetalias{PublicationI} we have shown 
that $\Pi^T$ gives the force-dipole at the object's
origin, induced by external forcing. The geometry of a uniform spheroid
is invariant under inversion symmetry, thus, the corresponding $\Pi$ tensor must vanish.}.
Hence, in the far field dynamics of two uniform spheroids the alignability and
direct interaction are absent, \ie, $\lambda=\zeta=0$ in
Eqs.~\eqref{eq:ellipsEOM1} and \eqref{eq:ellipsEOM2}. 
The resulting
picture, arising from Eqs.~\eqref{eq:ellipsEOM1} and
\eqref{eq:ellipsEOM2}, is that $\theta$ increases linearly with time
until saturating to a constant value, which depends on $R(t=0)$, and
the objects move away from each other with a constant terminal
velocity~\footnote{This behavior occurs in the limit of far-field
  dynamics. When the initial separation is small, the assumption
  $\theta \ll 1$ is not valid, or a periodic motion might appear; see
  Refs.~\cite{Kutteh2010,Kim1986,Kim1985}.}; see also dash-dotted
green curves in Fig.~\ref{fig:repulsion}.  Hence, the role of
hydrodynamic interactions in this case is solely the generation of
opposite tilts between the objects, which, as a result, move in
opposite directions.  In this case, Eq.~\eqref{eq:x} becomes equivalent to
the one-dimensional problem from classical mechanics of two particles
with a central repulsive potential, $\ddot{x}= A x^{-2}$, and the
initial conditions $x(0)=x_0$ and $\dot{x}(0)=0$.  The qualitative
behavior is apparent from the corresponding velocity equation
$\dot{x}=\sqrt{2 A(x^{-1}_0-x^{-1})}$.  The positive velocity
increases $x$, which in turn increases $\dot{x}$ toward a constant value, 
whereby $x$ continues to grow linearly with time.

The two power laws --- $R(t)\sim t^{1/3}$, derived theoretically for self-aligning
spheroids and demonstrated numerically for irregular objects, and 
$R(t)\sim t$, derived in the symmetric case of two uniform spheroids --- are universal.
Thus, the repulsive dynamics of pairs of symmetric objects 
and self-aligning objects are superficially similar, in that both arise from opposite tilts of the two objects.
However, the mechanisms of the two repulsions differ qualitatively.
The additional tendency of the latter objects to align with the force leads to a decrease in the relative
velocity, as reflected by a weaker power law for the increase of separation with time.
In addition, the terminal relative velocity in the case of two uniform spheroids is sensitive to 
the initial separation, as well as to the initial tilts. By contrast,
the self-aligning pair has a stable asymptotic velocity independent of the initial state of alignment
(assuming that $R(t=0)\gg l$ such that the far-field equations,
Eqs. \eqref{eq:EOM1}--\eqref{eq:EOM3}, are valid).

\subsubsection{Effect of longitudinal separation}
\label{sssec:longitudinal}

Up until now we have examined the simple case of constant force with
initial separation perpendicular to its direction. 
As explained below, a small additional separation along the direction of the force 
should not alter qualitatively the transversal repulsion.

The evolution of the longitudinal separation differs from the transversal one.    
The two components governing the far-field transversal dynamics --- mutual relative
orientation between the objects (the opposite tilt) and the direct
interaction which decays as $R^{-2}$ --- are weaker, or even absent,
in the longitudinal dynamics. First, the direct interaction, originating from the
off-diagonal blocks of the pair-mobility $\Mmat{A}$, vanishes
due to the constraint that $\Mmat{A}$ is a symmetric matrix~\cite{Condiff&Dahler1966,Happel&Brenner,Landau&Lifshitz}.
The components relating one object's linear velocity along the $\hat{z}$ direction
with forcing on the other, satisfy $\Mmat{A}^{ab}_{zz}=\Mmat{A}^{ba}_{zz}$,
which implies 
$\dot{R}_z=(\Mmat{A}^{ab}_{zz}-\Mmat{A}^{ba}_{zz})(-F)=0$. 
Hence, relative longitudinal dynamics is solely dictated by relative orientation between
the objects, which in the far-field regime corresponds to 
the difference between the (unperturbed) self-responses of each object.

When $R_z \ll R_{\perp}$ we can approximate the objects' orientations
by two opposite tilts of their eigen-directions. If the object's shape is invariant 
under rotations about the eigen-direction, e.g., self-aligning spheroid, 
such relative orientation cannot yield relative velocity along the $z$-axis.
In the case of arbitrarily shaped, self-aligning objects, we expect that
any asymmetry about the eigen-direction approximately averages out by the rotation
of each object. Thus, the effect of opposite tilts on the relative 
translation along the direction of the force is weaker than that on the transversal one.
Indeed, the examples in~\citetalias{PublicationI} showed that the longitudinal separation
evolves slowly in time and seems to saturate at long times.

\section{Discussion}
\label{sec:discussion}

This work, together with \citetalias{PublicationI}, aims to provide a comprehensive description
of the translational and orientational hydrodynamic interactions between 
two forced objects, focusing on the generic features of these interactions.
In particular, we have derived a formalism to predict from the system's symmetry 
whether the hydrodynamic interactions create relative motion (orientational or translational) between 
the objects. Where relative motion is present, we have analyzed its
multipole expansion. While our symmetry-based results for the instantaneous interaction are rigorous,
we could provide only qualitative general trends concerning the time-dependent relative motion,
focusing particularly on irregular, self-aligning objects, and on the asymptotic dynamics
at long times.

The present, second article has been devoted to 
the relative motion between two equally forced objects.  
The first part (Sec.~\ref{sec:Instantaneous}) has established 
the basic geometry dependence of the effective interaction.
We have proven that invariance
under spatial inversion precludes any instantaneous relative translation.
In the second part  (Sec.~\ref{sec:alignable}), we have treated the effect
of hydrodynamic interactions on time-dependent trajectories.
We have demonstrated how the characteristic $R_{\perp}(t)\sim t^{1/3}$
repulsion between two self-aligning objects
differs qualitatively from the counterpart, asymptotic $R_{\perp}(t)\sim t$ behavior,
which corresponds to two uniform spheroids. 
The preferred alignment of the individual objects reduces the relative tilt
as they get further apart, thus decreasing their relative translation
compared to the constantly tilted spheroids.
This case highlights the sharp contrast between motion conditioned by current configuration 
and time-integrated motion.
Two initially aligned spheroids do not have instantaneous relative velocity, while two
self-aligning objects do (due to the direct interaction, $\sim R^{-2}$, term).
Yet, the orientational interaction between the spheroids makes them 
oppositely tilt and achieve with time an asymptotic translational velocity which exceeds that of the 
self-aligning objects.

We have applied our general symmetry criterion to systems with
confining boundaries, which break inversion symmetry. This geometrical consideration,
without any further detail, accounts for the apparent interactions originated in hydrodynamic coupling,
which were observed in  optical-tweezers experiments involving two confined configurations~\cite{Larsen&Grier1997,Sokolov_etal2011}.
Previous works~\cite{Squires&Brenner2000,Sokolov_etal2011} examined these apparent interactions
for two point-like objects.
Our treatment shows that the existence of the effect can be inferred by symmetry.
It is present, therefore, in more general situations, such as non-spherical objects 
(e.g., Fig~\ref{fig:WallRing}(a)) and arbitrary separations, including objects in close proximity.

The general theory presented here can be used to obtain simple qualitative predictions,
which are readily testable in experiment.
A particularly simple example is the sedimentation of two identical spheres, positioned one above the other 
parallel to a vertical planar wall. This configuration will result in one sphere 
instantaneously approaching the wall, while the other being repelled from the wall.
Another example is a modification of the experiment presented in Ref.~\cite{Sokolov_etal2011}.
Two identical spheres placed in a ring and forced
in the radial direction will develop relative translation in the tangential direction.
We note that the general symmetry criterion concerns only the existence or
absence of interaction. To determine the sign of the interaction, whether it is repulsive or 
attractive, one needs additional information such as the Green's function of the 
given hydrodynamic problem.  

The symmetry arguments, which we have applied in the first part of the paper, can be found useful
in the examination of orientational dynamics.
In a system with spatial inversion symmetry, object $a$ has the same translational response 
as object $b$, and an opposite rotational response. Hence, when the objects are subjected to {\em opposite} forces,
e.g., in the presence of a central force of interaction between them, 
they must rotate in the {\em same} sense, and spatial inversion symmetry will be maintained at all times.
This means that, in the dynamics of an enantiomorphic pair under opposite drive, 
relative orientations, possessing a spatial inversion symmetry, are fixed points in the orientational space, preserving
a constant phase difference.
The study of this effect and its applications is postponed to a future publication.   

Another result, which can be verified in a simple experimental setup,
is the asymptotic power-law time dependence of the separation between two self-aligning objects.
For the particular system comprising a pair of self-aligning (non-uniform) prolate spheroids 
we have shown analytically a $t^{1/3}$ trend (Sec.~\ref{sec:alignable}).
In the case of two, arbitrarily shaped self-aligning objects, where repulsion 
is not a general law (see~\citetalias{PublicationI}), we have reported several examples with
a similar trend. These examples, which are represented by the red curve in Fig.~\ref{fig:repulsion},
are not sensitive to initial conditions. 
For spheroids, the effect of opposite tilts on the mutual repulsion is captured by the  
positive glide parameter $\alpha$ in Eq.~\eqref{eq:ellipsEOM1}.  
The sign of $\alpha$ is dictated by the elongated shape of the individual spheroid.
(An oblate spheroid has a negative $\alpha$.). 
Therefore, good candidates for arbitrarily shaped, repulsive pairs may be elongated objects, whose
properties, when averaged over rotations about the eigen-direction, resemble those of self-aligning spheroids. 
A suggestion for an experiment includes tracking the  
sedimentation of two micron-sized, self-aligning objects in a viscous fluid,
where optical traps can be used to place the objects at a fixed initial
separation, perpendicular to gravity.  
The effect should not depend on the initial orientations of the objects, 
as their alignability guarantees that after a short transient they will be close 
to their aligned state.

Finally, the distinction between irregular objects
and regular ones, on the level of a pair of objects, 
should be significant, in particular, in driven suspensions with many-body interactions. 
Traditionally, theories and simulations of fluidized beds focused on colloidal objects of spherical
or rod-like shape (see reviews in Refs.~\cite{Ramaswamy2001} and~\cite{Guazzelli&Hinch2011}
and references therein); however, 
the case of self-aligning irregular objects might give rise to new phenomena.  
For example,  
sedimentation of spheres involves only three-body effective interactions, 
whereas a suspension of sedimenting irregular objects will include effective {\em pair}-interactions.
Such pair interactions should affect the objects' velocity correlations,
as will be addressed in a future publication.
 
\begin{acknowledgments}
This research has been supported by the US--Israel Binational Science Foundation
(Grant no.\ 2012090).
\end{acknowledgments}

\appendix

\section{Notation}
\label{sec:notation}

The dynamics of arbitrarily shaped objects is complex and involves mathematical
structures of various dimensions. 
In order to facilitate the readability of the formalism, we use the following notation regarding vectors,
tensors, and matrices:
\begin{enumerate}

 \item 3-vectors are denoted by an arrow, $\vec{v}$, and unit 3-vectors
   by a hat, $\hat{v}$.

 \item Matrices are marked by a blackboard-bold letter, e.g.,
   $\Mmat{M}$, where the dimension of the matrix is understood from
   the context.

 \item Tensors of rank 3 are denoted by a capital Greek letter, e.g., $\Phi$.
 
 \item
 $\Mmat{I}_{n\times n}$ is the $n\times n$ identity matrix.

 \item
Tensor multiplication\,---\,the dot notation, $\cdot$\,---\,denotes a
contraction over one index. The double dot notation, $:$, denotes a
contraction over two indices. Thus, given a tensor $\Upsilon$ of rank
$N$ and a tensor $\Xi$ of rank $M>N$, the tensors $\Upsilon \cdot \Xi$
and $\Upsilon : \Xi$ are tensors of rank $N+M-2$ and $N+M-4$.
For example, for $\Upsilon$ of rank 2 and $\Xi$ of rank 3,
$(\Xi \cdot \Upsilon)_{ikj}=\Upsilon_{is} \, \Xi_{skj}$ and
$(\Upsilon : \Xi)_{j}=\Upsilon_{ks} \, \Xi_{skj}$.

\item
The matrix $\vec{Y}^\times$ obtained from the vector $\vec{Y}$ is
defined as $(\vec{Y}^\times)_{ij}=\epsilon_{ikj}Y_k$, such that, for
any vector $\vec{X}$,
$\vec{Y}^\times\cdot\vec{X}=\vec{Y}\times\vec{X}$.

\end{enumerate}

\section{Pair-Mobility Matrix of Two Spheres Near a Wall}
\label{app:Wall}

Here we provide more details regarding the pair-mobility matrix of a system
comprising two spheres near a wall.
We  write down the blocks structure
of the corresponding $\Mmat{A}$ and provide an explicit expressions for the case of point-like objects.
A schematic description of the system is given in Fig.~\ref{fig:WallRing}b.
Without loss of generality we assume that the spheres are
located along the $x$-axis, where $\vec{R}=x \hat{x}$ points from the origin of sphere $a$
to the origin of sphere $b$, and the wall is placed parallel to them at height $z=h$.
The spheres' radii are denoted by $\rho$. 
Hereafter we consider the projection of $\Mmat{A}$ onto the $xz$-plane. 
The properties of the $y$-axis components can be deduced from
the additional symmetry of reflection about the $xz$-plane, which was not
included in our analysis above. 
According to the discussion in Sec.~\ref{sec:WallRing} the pair-mobility matrix of this
system has the following form:
\begin{equation}
\Mmat{A}=
\begin{pmatrix}
  A^{\text{self}}_{xx}	& A^{\text{self}}_{x z} 
& A^{\text{coupling}}_{xx}	& A^{\text{coupling}}_{x z} \\
  A^{\text{self}}_{z x} & A^{\text{self}}_{zz} 
&  A^{\text{coupling}}_{z x} & A^{\text{coupling}}_{zz} \\
A^{\text{coupling}}_{xx}	& -A^{\text{coupling}}_{x z}
&  A^{\text{self}}_{xx}	& -A^{\text{self}}_{x z} \\ 
-A^{\text{coupling}}_{z x} & A^{\text{coupling}}_{zz}
& -A^{\text{self}}_{z x} & A^{\text{self}}_{zz} 
\end{pmatrix}.
\label{eq:FullWall}
\end{equation}
The number of independent components can be reduced further by using the fact
that $\Mmat{A}$ is symmetric. This property is not related to the system geometry,
which is the issue of Sec.~\ref{sec:WallRing}, but rather results from
Onsager relations or conservation of angular momentum
in the system~\cite{Condiff&Dahler1966,Happel&Brenner,Landau&Lifshitz}.  
The symmetry of $\Mmat{A}$ connects between the $xz$ and $zx$ components:
$A^{\text{self}}_{x z}=A^{\text{self}}_{z x}$ and 
$A^{\text{coupling}}_{x z}=-A^{\text{coupling}}_{z x}$.
Finally, we get
\begin{equation}
\Mmat{A}=
\begin{pmatrix}
  A^{\text{self}}_{xx}	& A^{\text{self}}_{x z} 
& A^{\text{coupling}}_{xx}	& A^{\text{coupling}}_{x z} \\
  A^{\text{self}}_{x z} & A^{\text{self}}_{zz} 
&  -A^{\text{coupling}}_{ x z} & A^{\text{coupling}}_{zz} \\
A^{\text{coupling}}_{xx}	& -A^{\text{coupling}}_{x z}
&  A^{\text{self}}_{xx}	& -A^{\text{self}}_{x z} \\ 
A^{\text{coupling}}_{x z} & A^{\text{coupling}}_{zz}
& -A^{\text{self}}_{x z} & A^{\text{self}}_{zz} 
\end{pmatrix}.
\label{eq:FullWall2}
\end{equation}

In the case of point-like objects, i.e., spheres with 
infinitely small radius, the blocks can be calculated
explicitly. 
The self blocks are given by the 
self-mobility of a single sphere near a plane wall 
(first-order in $\rho/h$)~\cite{Happel&Brenner}  
$$
\Mmat{A}^{\text{self}}=
\frac{1}{6\pi\eta \rho} 
\begin{pmatrix}
	1-\frac{9}{16}\frac{\rho}{h}
	&0 \\
	0  &
	1-\frac{9}{8}\frac{\rho}{h}
\end{pmatrix}.
$$
The coupling blocks, which correspond to the direct
hydrodynamic interaction between the spheres, are given by 
the Green function of the Stokes equation with a no-slip, plane wall boundary~\cite{Pozrikidis},
$$\Mmat{A}^{\text{coupling}}=
\frac{1}{8\pi\eta}
\begin{pmatrix}
	\frac{ 2x^2 (4 h^2 + x^2)^{5/2}-2(12 h^4 + 4 h^2 x^2 + x^4) |x|^3}
	{(4 h^2 + x^2)^{5/2} |x|^3} & 
	-\frac{12 h^3 x}{(4 h^2 + x^2)^{5/2}} \\
	\frac{12 h^3 x}{(4 h^2 + x^2)^{5/2}} & 
	\frac{(4 h^2 + x^2)^{5/2} - 48 h^4 |x| - 10 h^2 |x|^3 -|x|^5}
	{(4 h^2 + x^2)^{5/2} |x|}
\end{pmatrix}.
$$
The component $\Mmat{A}^{\text{coupling}}_{xz}=-12 h^3 x/(4 h^2 + x^2)^{5/2}$
is the one which was used in Ref.~\cite{Squires&Brenner2000} 
to explain the effective attraction between two like-charged spheres near a similarly 
charged wall.

\section{Spheroid parameters and integration scheme
for irregular objects}
\label{app:numerical}

Here we provide more details on the derivation of 
Eqs.~\eqref{eq:ellipsEOM1}-\eqref{eq:ellipsEOM2}, together with indicating
the specific parameters used for Fig.~\ref{fig:repulsion}.
In addition, we introduce the integration scheme for the far-field
dynamics of two irregular objects.

Eqs.~\eqref{eq:EOM1}-\eqref{eq:EOM3} contain the single-object-dependent tensors,
and the derivatives of the Oseen tensor, 
$\Mmat{G}_{ij}(\vec{R})=1/(8\pi R^2)(\delta_{ij}+R_i R_j/R^2)$. 
We calculate the object-dependent tensors of self-aligning spheroids as follows:
In the case of a uniform prolate spheroid, these tensors can be found explicitly,
using the results in Ref.~\cite{BrennerIV}. For example, when the major axis is parallel to
the $z$-axis we have $\Mmat{A}_{0,h=0}=diag(a_{\perp},a_{\perp},a_{\parallel})$, where the parameters
$a_{\perp}$ and $a_{\parallel}$ depend on the aspect ratio, and $\Mmat{T}_{0,h=0},\Pi_{0,h=0}=0$
(the components of $\Psi$ do not enter into Eq.\eqref{eq:ellipsEOM2}, see footnote~\cite{Note11}). 
Then, the properties of self-aligning prolate spheroids ($h\neq0$), 
$\Mmat{A}_{0,h}$, $\Mmat{T}_{0,h}$ and $\Pi_{0,h}$, are derived by change of origin transformation; 
see Ref.~\cite{Happel&Brenner} for transformations which correspond to the tensors of rank 2,
and Appendix C in Ref.~\cite{PublicationI}
for transformations concerning the tensors of rank 3. Eventually, for a tilted spheroid, the corresponding tensors are given 
by a rotation transformation.

The parameters $\alpha$, $\lambda$ and $\zeta$ in Eqs.~\eqref{eq:ellipsEOM1}-\eqref{eq:ellipsEOM2}
depend on the components of $\Mmat{A}_{0,h}$, $\Mmat{T}_{0,h}$ and $\Pi_{0,h}$, respectively.
In particular, $\lambda$ and $\zeta$ change linearly with $h$.
The trajectories presented in Fig.~\ref{fig:repulsion} correspond to spheroids with
aspect ratio of 4. The dash-dotted green and dotted blue curves, respectively, are solutions to 
Eqs.~\eqref{eq:ellipsEOM1}--\eqref{eq:ellipsEOM2} with $h=0$ (which gives $\alpha\approx 1$,
$\lambda,\,\zeta=0$) and $h\approx 0.31$ ( $\alpha\approx 0.6$,
$\lambda\approx1.25$, $\zeta\approx0.95$).

The dynamics of a symmetric system comprising two spheroids (Fig.~\ref{fig:spheroids}) can be described
by the reduced equations~\eqref{eq:ellipsEOM1}-\eqref{eq:ellipsEOM2} 
for one angle $\theta(t)$ and one-dimensional separation $x(t)$. However, in the general
case of two irregular objects we are compelled to integrate the full far-field equations~\eqref{eq:EOM1}-\eqref{eq:EOM3}.
Below we describe the details of the integration scheme.

The coordinate space includes the separation vector $\vec{R}$
and orientational parameters for each object, 
these are represented by Euler-Rodriguez 4-parameters (or unit quaternions), 
$(\Gamma^a,\vec{\Omega}^a)$ and $(\Gamma^b,\vec{\Omega}^b)$.
The tensorial properties of a given object, such as the matrix $\Mmat{A}$ or the tensor of
rank 3 $\Pi$, are calculated only once, in a reference frame affixed to the object.
For stokeslets object this properties can be derived self-consistently as described in 
subsection VA in \citetalias{PublicationI}.  

Knowing the properties in the body reference frame, e.g., $\Mmat{A}^{0}$ or $\Pi^{0}$,
one can use a rotation transformation to calculate them in any orientation
$(\Gamma,\vec{\Omega})$: 
$$
\Mmat{A}_{ij}(\Gamma,\vec{\Omega})=\Mmat{R}_{il}(\Gamma,\vec{\Omega}) 
\Mmat{A}^0_{lm} \Mmat{R}^T_{mj}(\Gamma,\vec{\Omega}),
$$ 
$$
\Pi_{ijk}(\Gamma,\vec{\Omega})=\Mmat{R}_{im}(\Gamma,\vec{\Omega}) 
\Pi^0_{mls} \Mmat{R}^T_{lj}(\Gamma,\vec{\Omega}) \Mmat{R}^T_{sk}(\Gamma,\vec{\Omega}),
$$
where 
$$\Mmat{R}_{ij}(\Gamma,\vec{\Omega})=
(1-2\Omega^2)\delta_{ij}+2\Gamma\epsilon_{ikj}\Omega_k +2\Omega_i\Omega_j$$
is the rotation matrix which is a polynomial in the orientational parameters.

Finally, the equations for the evolution of 
$\left(\vec{R},\Gamma^a,\vec{\Omega}^a,\Gamma^b,\vec{\Omega}^b\right)$
can be written using Eq.~\eqref{eq:EOM1},
and Eq.~\eqref{eq:EOM2}--\eqref{eq:EOM3} together with the linear relation
between angular velocity and time derivative of the orientation parameters 
	\begin{equation}
		\begin{pmatrix}
		\dot{\Gamma} \\
		\dot{\vec{\Omega}}
	\end{pmatrix}=
	\frac{1}{2}
	\begin{pmatrix}
		0  & -\vec{\omega}^T\\
		\vec{\omega} &\vec{\omega}^{\times}
	\end{pmatrix}
	\begin{pmatrix}
		\Gamma \\
		\vec{\Omega}
	\end{pmatrix},
\label{eq:angular}
	\end{equation}
	where $\vec{\omega}^{\times}_{ij}=\epsilon_{ikj}\omega_k$.

\bibliography{pair_interactionII_re160502}

\begin{thebibliography}{42}%
\makeatletter
\providecommand \@ifxundefined [1]{%
 \@ifx{#1\undefined}
}%
\providecommand \@ifnum [1]{%
 \ifnum #1\expandafter \@firstoftwo
 \else \expandafter \@secondoftwo
 \fi
}%
\providecommand \@ifx [1]{%
 \ifx #1\expandafter \@firstoftwo
 \else \expandafter \@secondoftwo
 \fi
}%
\providecommand \natexlab [1]{#1}%
\providecommand \enquote  [1]{``#1''}%
\providecommand \bibnamefont  [1]{#1}%
\providecommand \bibfnamefont [1]{#1}%
\providecommand \citenamefont [1]{#1}%
\providecommand \href@noop [0]{\@secondoftwo}%
\providecommand \href [0]{\begingroup \@sanitize@url \@href}%
\providecommand \@href[1]{\@@startlink{#1}\@@href}%
\providecommand \@@href[1]{\endgroup#1\@@endlink}%
\providecommand \@sanitize@url [0]{\catcode `\\12\catcode `\$12\catcode
  `\&12\catcode `\#12\catcode `\^12\catcode `\_12\catcode `\%12\relax}%
\providecommand \@@startlink[1]{}%
\providecommand \@@endlink[0]{}%
\providecommand \url  [0]{\begingroup\@sanitize@url \@url }%
\providecommand \@url [1]{\endgroup\@href {#1}{\urlprefix }}%
\providecommand \urlprefix  [0]{URL }%
\providecommand \Eprint [0]{\href }%
\providecommand \doibase [0]{http://dx.doi.org/}%
\providecommand \selectlanguage [0]{\@gobble}%
\providecommand \bibinfo  [0]{\@secondoftwo}%
\providecommand \bibfield  [0]{\@secondoftwo}%
\providecommand \translation [1]{[#1]}%
\providecommand \BibitemOpen [0]{}%
\providecommand \bibitemStop [0]{}%
\providecommand \bibitemNoStop [0]{.\EOS\space}%
\providecommand \EOS [0]{\spacefactor3000\relax}%
\providecommand \BibitemShut  [1]{\csname bibitem#1\endcsname}%
\let\auto@bib@innerbib\@empty
\bibitem [{\citenamefont {Happel}\ and\ \citenamefont
  {Brenner}(1983)}]{Happel&Brenner}%
  \BibitemOpen
  \bibfield  {author} {\bibinfo {author} {\bibfnamefont {J.}~\bibnamefont
  {Happel}}\ and\ \bibinfo {author} {\bibfnamefont {H.}~\bibnamefont
  {Brenner}},\ }\href@noop {} {\emph {\bibinfo {title} {Low Reynolds Number
  Hydrodynamics: with Special Applications to Particulate Media}}}\ (\bibinfo
  {publisher} {Martinus Nijhoff, The Hague},\ \bibinfo {year}
  {1983})\BibitemShut {NoStop}%
\bibitem [{\citenamefont {Russel}\ \emph {et~al.}(1989)\citenamefont {Russel},
  \citenamefont {Saville},\ and\ \citenamefont {Schowalter}}]{Russel}%
  \BibitemOpen
  \bibfield  {author} {\bibinfo {author} {\bibfnamefont {W.~B.}\ \bibnamefont
  {Russel}}, \bibinfo {author} {\bibfnamefont {D.~A.}\ \bibnamefont {Saville}},
  \ and\ \bibinfo {author} {\bibfnamefont {W.~R.}\ \bibnamefont {Schowalter}},\
  }\href@noop {} {\emph {\bibinfo {title} {Colloidal Dispersions}}}\ (\bibinfo
  {publisher} {Cambridge University Press},\ \bibinfo {year}
  {1989})\BibitemShut {NoStop}%
\bibitem [{\citenamefont {Goldfriend}\ \emph {et~al.}(2015)\citenamefont
  {Goldfriend}, \citenamefont {Diamant},\ and\ \citenamefont
  {Witten}}]{PublicationI}%
  \BibitemOpen
  \bibfield  {author} {\bibinfo {author} {\bibfnamefont {T.}~\bibnamefont
  {Goldfriend}}, \bibinfo {author} {\bibfnamefont {H.}~\bibnamefont {Diamant}},
  \ and\ \bibinfo {author} {\bibfnamefont {T.~A.}\ \bibnamefont {Witten}},\
  }\href {\doibase 10.1063/1.4936894} {\bibfield  {journal} {\bibinfo
  {journal} {Phys. Fluids}\ }\textbf {\bibinfo {volume} {27}},\ \bibinfo
  {pages} {123303} (\bibinfo {year} {2015})}\BibitemShut {NoStop}%
\bibitem [{\citenamefont {Ramaswamy}(2001)}]{Ramaswamy2001}%
  \BibitemOpen
  \bibfield  {author} {\bibinfo {author} {\bibfnamefont {S.}~\bibnamefont
  {Ramaswamy}},\ }\href {\doibase 10.1080/00018730110050617} {\bibfield
  {journal} {\bibinfo  {journal} {Adv. Phys.}\ }\textbf {\bibinfo {volume}
  {50}},\ \bibinfo {pages} {297} (\bibinfo {year} {2001})}\BibitemShut
  {NoStop}%
\bibitem [{\citenamefont {Squires}\ and\ \citenamefont
  {Brenner}(2000)}]{Squires&Brenner2000}%
  \BibitemOpen
  \bibfield  {author} {\bibinfo {author} {\bibfnamefont {T.~M.}\ \bibnamefont
  {Squires}}\ and\ \bibinfo {author} {\bibfnamefont {M.~P.}\ \bibnamefont
  {Brenner}},\ }\href {\doibase 10.1103/PhysRevLett.85.4976} {\bibfield
  {journal} {\bibinfo  {journal} {Phys. Rev. Lett.}\ }\textbf {\bibinfo
  {volume} {85}},\ \bibinfo {pages} {4976} (\bibinfo {year}
  {2000})}\BibitemShut {NoStop}%
\bibitem [{\citenamefont {Larsen}\ and\ \citenamefont
  {Grier}(1997)}]{Larsen&Grier1997}%
  \BibitemOpen
  \bibfield  {author} {\bibinfo {author} {\bibfnamefont {A.~E.}\ \bibnamefont
  {Larsen}}\ and\ \bibinfo {author} {\bibfnamefont {D.~G.}\ \bibnamefont
  {Grier}},\ }\href {\doibase 10.1038/385230a0} {\bibfield  {journal} {\bibinfo
   {journal} {Nature}\ }\textbf {\bibinfo {volume} {385}},\ \bibinfo {pages}
  {230} (\bibinfo {year} {1997})}\BibitemShut {NoStop}%
\bibitem [{\citenamefont {Squires}(2001)}]{Squires2001}%
  \BibitemOpen
  \bibfield  {author} {\bibinfo {author} {\bibfnamefont {T.~M.}\ \bibnamefont
  {Squires}},\ }\href {\doibase 10.1017/S0022112001005432} {\bibfield
  {journal} {\bibinfo  {journal} {J. Fluid Mech.}\ }\textbf {\bibinfo {volume}
  {443}},\ \bibinfo {pages} {403} (\bibinfo {year} {2001})}\BibitemShut
  {NoStop}%
\bibitem [{\citenamefont {Sokolov}\ \emph {et~al.}(2011)\citenamefont
  {Sokolov}, \citenamefont {Frydel}, \citenamefont {Grier}, \citenamefont
  {Diamant},\ and\ \citenamefont {Roichman}}]{Sokolov_etal2011}%
  \BibitemOpen
  \bibfield  {author} {\bibinfo {author} {\bibfnamefont {Y.}~\bibnamefont
  {Sokolov}}, \bibinfo {author} {\bibfnamefont {D.}~\bibnamefont {Frydel}},
  \bibinfo {author} {\bibfnamefont {D.~G.}\ \bibnamefont {Grier}}, \bibinfo
  {author} {\bibfnamefont {H.}~\bibnamefont {Diamant}}, \ and\ \bibinfo
  {author} {\bibfnamefont {Y.}~\bibnamefont {Roichman}},\ }\href {\doibase
  10.1103/PhysRevLett.107.158302} {\bibfield  {journal} {\bibinfo  {journal}
  {Phys. Rev. Lett.}\ }\textbf {\bibinfo {volume} {107}},\ \bibinfo {pages}
  {158302} (\bibinfo {year} {2011})}\BibitemShut {NoStop}%
\bibitem [{\citenamefont {Nagar}\ and\ \citenamefont
  {Roichman}(2014)}]{Nagar&Roichman2014}%
  \BibitemOpen
  \bibfield  {author} {\bibinfo {author} {\bibfnamefont {H.}~\bibnamefont
  {Nagar}}\ and\ \bibinfo {author} {\bibfnamefont {Y.}~\bibnamefont
  {Roichman}},\ }\href {\doibase 10.1103/PhysRevE.90.042302} {\bibfield
  {journal} {\bibinfo  {journal} {Phys. Rev. E}\ }\textbf {\bibinfo {volume}
  {90}},\ \bibinfo {pages} {042302} (\bibinfo {year} {2014})}\BibitemShut
  {NoStop}%
\bibitem [{\citenamefont {Beatus}\ \emph {et~al.}(2012)\citenamefont {Beatus},
  \citenamefont {Bar-Ziv},\ and\ \citenamefont {Tlusty}}]{Beatus_etal2012}%
  \BibitemOpen
  \bibfield  {author} {\bibinfo {author} {\bibfnamefont {T.}~\bibnamefont
  {Beatus}}, \bibinfo {author} {\bibfnamefont {R.~H.}\ \bibnamefont {Bar-Ziv}},
  \ and\ \bibinfo {author} {\bibfnamefont {T.}~\bibnamefont {Tlusty}},\ }\href
  {\doibase http://dx.doi.org/10.1016/j.physrep.2012.02.003} {\bibfield
  {journal} {\bibinfo  {journal} {Phys. Rep.}\ }\textbf {\bibinfo {volume}
  {516}},\ \bibinfo {pages} {103} (\bibinfo {year} {2012})}\BibitemShut
  {NoStop}%
\bibitem [{\citenamefont {Bretherton}(1962)}]{Bretherton1962}%
  \BibitemOpen
  \bibfield  {author} {\bibinfo {author} {\bibfnamefont {F.~P.}\ \bibnamefont
  {Bretherton}},\ }\href {\doibase 10.1017/S002211206200124X} {\bibfield
  {journal} {\bibinfo  {journal} {J. Fluid Mech.}\ }\textbf {\bibinfo {volume}
  {14}},\ \bibinfo {pages} {284} (\bibinfo {year} {1962})}\BibitemShut
  {NoStop}%
\bibitem [{\citenamefont {Makino}\ and\ \citenamefont
  {Doi}(2005)}]{Makino&Doi2005}%
  \BibitemOpen
  \bibfield  {author} {\bibinfo {author} {\bibfnamefont {M.}~\bibnamefont
  {Makino}}\ and\ \bibinfo {author} {\bibfnamefont {M.}~\bibnamefont {Doi}},\
  }\href {\doibase http://dx.doi.org/10.1063/1.2107867} {\bibfield  {journal}
  {\bibinfo  {journal} {Phys. Fluids}\ }\textbf {\bibinfo {volume} {17}},\
  \bibinfo {pages} {103605} (\bibinfo {year} {2005})}\BibitemShut {NoStop}%
\bibitem [{\citenamefont {Purcell}(1977)}]{Purcell1977}%
  \BibitemOpen
  \bibfield  {author} {\bibinfo {author} {\bibfnamefont {E.~M.}\ \bibnamefont
  {Purcell}},\ }\href@noop {} {\bibfield  {journal} {\bibinfo  {journal} {Am.
  J. Phys}\ }\textbf {\bibinfo {volume} {45}},\ \bibinfo {pages} {3} (\bibinfo
  {year} {1977})}\BibitemShut {NoStop}%
\bibitem [{\citenamefont {Brenner}(1964{\natexlab{a}})}]{BrennerII}%
  \BibitemOpen
  \bibfield  {author} {\bibinfo {author} {\bibfnamefont {H.}~\bibnamefont
  {Brenner}},\ }\href {\doibase http://dx.doi.org/10.1016/0009-2509(64)85051-X}
  {\bibfield  {journal} {\bibinfo  {journal} {Chem. Eng. Sci.}\ }\textbf
  {\bibinfo {volume} {19}},\ \bibinfo {pages} {599} (\bibinfo {year}
  {1964}{\natexlab{a}})}\BibitemShut {NoStop}%
\bibitem [{\citenamefont {Brenner}\ and\ \citenamefont
  {O{'}Neill}(1972)}]{Brenner&Oneill1972}%
  \BibitemOpen
  \bibfield  {author} {\bibinfo {author} {\bibfnamefont {H.}~\bibnamefont
  {Brenner}}\ and\ \bibinfo {author} {\bibfnamefont {M.~E.}\ \bibnamefont
  {O{'}Neill}},\ }\href {\doibase
  http://dx.doi.org/10.1016/0009-2509(72)85029-2} {\bibfield  {journal}
  {\bibinfo  {journal} {Chem. Eng. Sci.}\ }\textbf {\bibinfo {volume} {27}},\
  \bibinfo {pages} {1421} (\bibinfo {year} {1972})}\BibitemShut {NoStop}%
\bibitem [{\citenamefont {Kim}\ and\ \citenamefont
  {Karrila}(2005)}]{Kim&Karrila}%
  \BibitemOpen
  \bibfield  {author} {\bibinfo {author} {\bibfnamefont {S.}~\bibnamefont
  {Kim}}\ and\ \bibinfo {author} {\bibfnamefont {S.~J.}\ \bibnamefont
  {Karrila}},\ }\href@noop {} {\emph {\bibinfo {title} {Microhydrodynamics:
  Principles and Selected Applications}}}\ (\bibinfo  {publisher} {Dover
  Publications},\ \bibinfo {year} {2005})\BibitemShut {NoStop}%
\bibitem [{\citenamefont {Goldman}\ \emph {et~al.}(1966)\citenamefont
  {Goldman}, \citenamefont {Cox},\ and\ \citenamefont
  {Brenner}}]{Goldman_etal1966}%
  \BibitemOpen
  \bibfield  {author} {\bibinfo {author} {\bibfnamefont {A.}~\bibnamefont
  {Goldman}}, \bibinfo {author} {\bibfnamefont {R.}~\bibnamefont {Cox}}, \ and\
  \bibinfo {author} {\bibfnamefont {H.}~\bibnamefont {Brenner}},\ }\href
  {\doibase http://dx.doi.org/10.1016/0009-2509(66)85036-4} {\bibfield
  {journal} {\bibinfo  {journal} {Chem. Eng. Sci.}\ }\textbf {\bibinfo {volume}
  {21}},\ \bibinfo {pages} {1151} (\bibinfo {year} {1966})}\BibitemShut
  {NoStop}%
\bibitem [{\citenamefont {Wakiya}(1965)}]{Wakiya1965}%
  \BibitemOpen
  \bibfield  {author} {\bibinfo {author} {\bibfnamefont {S.}~\bibnamefont
  {Wakiya}},\ }\href {\doibase 10.1143/JPSJ.20.1502} {\bibfield  {journal}
  {\bibinfo  {journal} {J. Phys. Soc. Jpn.}\ }\textbf {\bibinfo {volume}
  {20}},\ \bibinfo {pages} {1502} (\bibinfo {year} {1965})}\BibitemShut
  {NoStop}%
\bibitem [{\citenamefont {Felderhof}(1977)}]{Felderhof1977}%
  \BibitemOpen
  \bibfield  {author} {\bibinfo {author} {\bibfnamefont {B.}~\bibnamefont
  {Felderhof}},\ }\href {\doibase
  http://dx.doi.org/10.1016/0378-4371(77)90111-X} {\bibfield  {journal}
  {\bibinfo  {journal} {Physica A}\ }\textbf {\bibinfo {volume} {89}},\
  \bibinfo {pages} {373} (\bibinfo {year} {1977})}\BibitemShut {NoStop}%
\bibitem [{\citenamefont {Jeffrey}\ and\ \citenamefont
  {Onishi}(1984)}]{Jeffrey&Onishi1984}%
  \BibitemOpen
  \bibfield  {author} {\bibinfo {author} {\bibfnamefont {D.~J.}\ \bibnamefont
  {Jeffrey}}\ and\ \bibinfo {author} {\bibfnamefont {Y.}~\bibnamefont
  {Onishi}},\ }\href {\doibase 10.1017/S0022112084000355} {\bibfield  {journal}
  {\bibinfo  {journal} {J. Fluid Mech.}\ }\textbf {\bibinfo {volume} {139}},\
  \bibinfo {pages} {261} (\bibinfo {year} {1984})}\BibitemShut {NoStop}%
\bibitem [{\citenamefont {Liao}\ and\ \citenamefont
  {Krueger}(1980)}]{Liao&Krueger1980}%
  \BibitemOpen
  \bibfield  {author} {\bibinfo {author} {\bibfnamefont {W.}~\bibnamefont
  {Liao}}\ and\ \bibinfo {author} {\bibfnamefont {D.~A.}\ \bibnamefont
  {Krueger}},\ }\href {\doibase 10.1017/S002211208000208X} {\bibfield
  {journal} {\bibinfo  {journal} {J. Fluid Mech.}\ }\textbf {\bibinfo {volume}
  {96}},\ \bibinfo {pages} {223} (\bibinfo {year} {1980})}\BibitemShut
  {NoStop}%
\bibitem [{\citenamefont {Kim}(1985)}]{Kim1985}%
  \BibitemOpen
  \bibfield  {author} {\bibinfo {author} {\bibfnamefont {S.}~\bibnamefont
  {Kim}},\ }\href {\doibase http://dx.doi.org/10.1016/0301-9322(85)90087-4}
  {\bibfield  {journal} {\bibinfo  {journal} {Int. J. Multiphase Flow}\
  }\textbf {\bibinfo {volume} {11}},\ \bibinfo {pages} {699} (\bibinfo {year}
  {1985})}\BibitemShut {NoStop}%
\bibitem [{\citenamefont {Kim}(1986)}]{Kim1986}%
  \BibitemOpen
  \bibfield  {author} {\bibinfo {author} {\bibfnamefont {S.}~\bibnamefont
  {Kim}},\ }\href {\doibase http://dx.doi.org/10.1016/0301-9322(86)90019-4}
  {\bibfield  {journal} {\bibinfo  {journal} {Int. J. Multiphase Flow}\
  }\textbf {\bibinfo {volume} {12}},\ \bibinfo {pages} {469} (\bibinfo {year}
  {1986})}\BibitemShut {NoStop}%
\bibitem [{\citenamefont {Karrila}\ \emph {et~al.}(1989)\citenamefont
  {Karrila}, \citenamefont {Fuentes},\ and\ \citenamefont
  {Kim}}]{Karrila_etal1989}%
  \BibitemOpen
  \bibfield  {author} {\bibinfo {author} {\bibfnamefont {S.~J.}\ \bibnamefont
  {Karrila}}, \bibinfo {author} {\bibfnamefont {Y.~O.}\ \bibnamefont
  {Fuentes}}, \ and\ \bibinfo {author} {\bibfnamefont {S.}~\bibnamefont
  {Kim}},\ }\href {\doibase http://dx.doi.org/10.1122/1.550040} {\bibfield
  {journal} {\bibinfo  {journal} {J. Rheol.}\ }\textbf {\bibinfo {volume}
  {33}},\ \bibinfo {pages} {913} (\bibinfo {year} {1989})}\BibitemShut
  {NoStop}%
\bibitem [{\citenamefont {Tran{-}Cong}\ and\ \citenamefont
  {Phan{-}Thien}(1989)}]{Cong&Thien1989}%
  \BibitemOpen
  \bibfield  {author} {\bibinfo {author} {\bibfnamefont {T.}~\bibnamefont
  {Tran{-}Cong}}\ and\ \bibinfo {author} {\bibfnamefont {N.}~\bibnamefont
  {Phan{-}Thien}},\ }\href {\doibase http://dx.doi.org/10.1063/1.857414}
  {\bibfield  {journal} {\bibinfo  {journal} {Phys. Fluids}\ }\textbf {\bibinfo
  {volume} {1}},\ \bibinfo {pages} {453} (\bibinfo {year} {1989})}\BibitemShut
  {NoStop}%
\bibitem [{\citenamefont {Carrasco}\ and\ \citenamefont {de~la
  Torre}(1999)}]{Carrasco&Torre1999}%
  \BibitemOpen
  \bibfield  {author} {\bibinfo {author} {\bibfnamefont {B.}~\bibnamefont
  {Carrasco}}\ and\ \bibinfo {author} {\bibfnamefont {J.~G.}\ \bibnamefont
  {de~la Torre}},\ }\href {\doibase
  http://dx.doi.org/10.1016/S0006-3495(99)77457-6} {\bibfield  {journal}
  {\bibinfo  {journal} {Biophys. J.}\ }\textbf {\bibinfo {volume} {76}},\
  \bibinfo {pages} {3044} (\bibinfo {year} {1999})}\BibitemShut {NoStop}%
\bibitem [{\citenamefont {Kutteh}(2010)}]{Kutteh2010}%
  \BibitemOpen
  \bibfield  {author} {\bibinfo {author} {\bibfnamefont {R.}~\bibnamefont
  {Kutteh}},\ }\href@noop {} {\bibfield  {journal} {\bibinfo  {journal} {J.
  Chem. Phys.}\ }\textbf {\bibinfo {volume} {132}} (\bibinfo {year}
  {2010})}\BibitemShut {NoStop}%
\bibitem [{\citenamefont {Cichocki}\ \emph {et~al.}(1994)\citenamefont
  {Cichocki}, \citenamefont {Felderhof}, \citenamefont {Hinsen}, \citenamefont
  {Wajnryb},\ and\ \citenamefont {B{\l}awzdziewicz}}]{Cichocki_etal1994}%
  \BibitemOpen
  \bibfield  {author} {\bibinfo {author} {\bibfnamefont {B.}~\bibnamefont
  {Cichocki}}, \bibinfo {author} {\bibfnamefont {B.~U.}\ \bibnamefont
  {Felderhof}}, \bibinfo {author} {\bibfnamefont {K.}~\bibnamefont {Hinsen}},
  \bibinfo {author} {\bibfnamefont {E.}~\bibnamefont {Wajnryb}}, \ and\
  \bibinfo {author} {\bibfnamefont {J.}~\bibnamefont {B{\l}awzdziewicz}},\
  }\href {\doibase http://dx.doi.org/10.1063/1.466366} {\bibfield  {journal}
  {\bibinfo  {journal} {J. Chem. Phys.}\ }\textbf {\bibinfo {volume} {100}},\
  \bibinfo {pages} {3780} (\bibinfo {year} {1994})}\BibitemShut {NoStop}%
\bibitem [{Note1()}]{Note1}%
  \BibitemOpen
  \bibinfo {note} {In \protect \citetalias {PublicationI} they were referred to
  as ``axially alignbale'' objects.}\BibitemShut {Stop}%
\bibitem [{\citenamefont {Krapf}\ \emph {et~al.}(2009)\citenamefont {Krapf},
  \citenamefont {Witten},\ and\ \citenamefont {Keim}}]{Krapf_etal2009}%
  \BibitemOpen
  \bibfield  {author} {\bibinfo {author} {\bibfnamefont {N.~W.}\ \bibnamefont
  {Krapf}}, \bibinfo {author} {\bibfnamefont {T.~A.}\ \bibnamefont {Witten}}, \
  and\ \bibinfo {author} {\bibfnamefont {N.~C.}\ \bibnamefont {Keim}},\ }\href
  {\doibase 10.1103/PhysRevE.79.056307} {\bibfield  {journal} {\bibinfo
  {journal} {Phys. Rev. E}\ }\textbf {\bibinfo {volume} {79}},\ \bibinfo
  {pages} {056307} (\bibinfo {year} {2009})}\BibitemShut {NoStop}%
\bibitem [{\citenamefont {Moths}\ and\ \citenamefont
  {Witten}(2013{\natexlab{a}})}]{Moths&Witten2013}%
  \BibitemOpen
  \bibfield  {author} {\bibinfo {author} {\bibfnamefont {B.}~\bibnamefont
  {Moths}}\ and\ \bibinfo {author} {\bibfnamefont {T.~A.}\ \bibnamefont
  {Witten}},\ }\href {\doibase 10.1103/PhysRevLett.110.028301} {\bibfield
  {journal} {\bibinfo  {journal} {Phys. Rev. Lett.}\ }\textbf {\bibinfo
  {volume} {110}},\ \bibinfo {pages} {028301} (\bibinfo {year}
  {2013}{\natexlab{a}})}\BibitemShut {NoStop}%
\bibitem [{\citenamefont {Moths}\ and\ \citenamefont
  {Witten}(2013{\natexlab{b}})}]{Moths&Witten2013b}%
  \BibitemOpen
  \bibfield  {author} {\bibinfo {author} {\bibfnamefont {B.}~\bibnamefont
  {Moths}}\ and\ \bibinfo {author} {\bibfnamefont {T.~A.}\ \bibnamefont
  {Witten}},\ }\href {\doibase 10.1103/PhysRevE.88.022307} {\bibfield
  {journal} {\bibinfo  {journal} {Phys. Rev. E}\ }\textbf {\bibinfo {volume}
  {88}},\ \bibinfo {pages} {022307} (\bibinfo {year}
  {2013}{\natexlab{b}})}\BibitemShut {NoStop}%
\bibitem [{\citenamefont {Condiff}\ and\ \citenamefont
  {Dahler}(1966)}]{Condiff&Dahler1966}%
  \BibitemOpen
  \bibfield  {author} {\bibinfo {author} {\bibfnamefont {D.~W.}\ \bibnamefont
  {Condiff}}\ and\ \bibinfo {author} {\bibfnamefont {J.~S.}\ \bibnamefont
  {Dahler}},\ }\href {\doibase http://dx.doi.org/10.1063/1.1726561} {\bibfield
  {journal} {\bibinfo  {journal} {J. Chem. Phys.}\ }\textbf {\bibinfo {volume}
  {44}},\ \bibinfo {pages} {3988} (\bibinfo {year} {1966})}\BibitemShut
  {NoStop}%
\bibitem [{\citenamefont {Landau}\ and\ \citenamefont
  {Lifshitz}(1980)}]{Landau&Lifshitz}%
  \BibitemOpen
  \bibfield  {author} {\bibinfo {author} {\bibfnamefont {L.}~\bibnamefont
  {Landau}}\ and\ \bibinfo {author} {\bibfnamefont {E.}~\bibnamefont
  {Lifshitz}},\ }\href@noop {} {\emph {\bibinfo {title} {Statistical Physics,
  Part 1, 3rd edition}}}\ (\bibinfo  {publisher} {Pergamon Press},\ \bibinfo
  {year} {1980})\BibitemShut {NoStop}%
\bibitem [{\citenamefont {Gonzalez}\ \emph {et~al.}(2004)\citenamefont
  {Gonzalez}, \citenamefont {Graf},\ and\ \citenamefont
  {Maddocks}}]{Gonzalez_etal2004}%
  \BibitemOpen
  \bibfield  {author} {\bibinfo {author} {\bibfnamefont {O.}~\bibnamefont
  {Gonzalez}}, \bibinfo {author} {\bibfnamefont {A.~B.~A.}\ \bibnamefont
  {Graf}}, \ and\ \bibinfo {author} {\bibfnamefont {J.~H.}\ \bibnamefont
  {Maddocks}},\ }\href {\doibase 10.1017/S0022112004001284} {\bibfield
  {journal} {\bibinfo  {journal} {J. Fluid Mech.}\ }\textbf {\bibinfo {volume}
  {519}},\ \bibinfo {pages} {133} (\bibinfo {year} {2004})}\BibitemShut
  {NoStop}%
\bibitem [{Note11()}]{Note11}%
  \BibitemOpen
  \bibinfo {note} {The object-dependent tensor $\Psi $, which appears in
  Eqs.~\protect \textup {\hbox {\mathsurround \z@ \protect \normalfont
  (\ignorespaces \ref {eq:EOM2}\unskip \@@italiccorr )}} and \protect \textup
  {\hbox {\mathsurround \z@ \protect \normalfont (\ignorespaces \ref
  {eq:EOM3}\unskip \@@italiccorr )}}, characterizes the response to a spatially
  symmetric flow gradient only (see Appendix B in \protect \citetalias
  {PublicationI}). When $\protect \mathaccentV {vec}17E{R}\perp \protect
  \mathaccentV {vec}17E{F}$, the flow gradients created by the force monopole
  on each object, $\pm \protect \mathaccentV {vec}17E{\nabla }\protect \mathbb
  {G}(\protect \mathaccentV {vec}17E{R})\cdot \protect \mathaccentV
  {vec}17E{F}$, are antisymmetric. Thus, the object-dependent terms vanish, and
  the angular velocities are solely affected by the vorticity $\sim \protect
  \mathaccentV {hat}05E{R}\times \protect \mathaccentV
  {vec}17E{F}/R^2$.}\BibitemShut {Stop}%
\bibitem [{\citenamefont {Favro}(1960)}]{Favro1960}%
  \BibitemOpen
  \bibfield  {author} {\bibinfo {author} {\bibfnamefont {L.~D.}\ \bibnamefont
  {Favro}},\ }\href {\doibase 10.1103/PhysRev.119.53} {\bibfield  {journal}
  {\bibinfo  {journal} {Phys. Rev.}\ }\textbf {\bibinfo {volume} {119}},\
  \bibinfo {pages} {53} (\bibinfo {year} {1960})}\BibitemShut {NoStop}%
\bibitem [{Note12()}]{Note12}%
  \BibitemOpen
  \bibinfo {note} {In \protect \citetalias {PublicationI} we have shown that
  $\Pi ^T$ gives the force-dipole at the object's origin, induced by external
  forcing. The geometry of a uniform spheroid is invariant under inversion
  symmetry, thus, the corresponding $\Pi $ tensor must vanish.}\BibitemShut
  {Stop}%
\bibitem [{Note13()}]{Note13}%
  \BibitemOpen
  \bibinfo {note} {This behavior occurs in the limit of far-field dynamics.
  When the initial separation is small, the assumption $\theta \ll 1$ is not
  valid, or a periodic motion might appear; see Refs.~\cite
  {Kutteh2010,Kim1986,Kim1985}.}\BibitemShut {Stop}%
\bibitem [{\citenamefont {Guazzelli}\ and\ \citenamefont
  {Hinch}(2011)}]{Guazzelli&Hinch2011}%
  \BibitemOpen
  \bibfield  {author} {\bibinfo {author} {\bibfnamefont {{\'E}.}~\bibnamefont
  {Guazzelli}}\ and\ \bibinfo {author} {\bibfnamefont {J.}~\bibnamefont
  {Hinch}},\ }\href {\doibase 10.1146/annurev-fluid-122109-160736} {\bibfield
  {journal} {\bibinfo  {journal} {Annu. Rev. Fluid Mech.}\ }\textbf {\bibinfo
  {volume} {43}},\ \bibinfo {pages} {97} (\bibinfo {year} {2011})}\BibitemShut
  {NoStop}%
\bibitem [{\citenamefont {Pozrikidis}(1992)}]{Pozrikidis}%
  \BibitemOpen
  \bibfield  {author} {\bibinfo {author} {\bibfnamefont {C.}~\bibnamefont
  {Pozrikidis}},\ }\href@noop {} {\emph {\bibinfo {title} {Boundary integral
  and singularity methods for linearized viscous flow}}}\ (\bibinfo
  {publisher} {Cambridge University Press},\ \bibinfo {year}
  {1992})\BibitemShut {NoStop}%
\bibitem [{\citenamefont {Brenner}(1964{\natexlab{b}})}]{BrennerIV}%
  \BibitemOpen
  \bibfield  {author} {\bibinfo {author} {\bibfnamefont {H.}~\bibnamefont
  {Brenner}},\ }\href {\doibase 10.1016/0009-2509(64)85084-3} {\bibfield
  {journal} {\bibinfo  {journal} {Chem. Eng. Sci.}\ }\textbf {\bibinfo {volume}
  {19}},\ \bibinfo {pages} {703} (\bibinfo {year}
  {1964}{\natexlab{b}})}\BibitemShut {NoStop}%
\end{thebibliography}%

\end{document}